\documentclass{aa}

\usepackage{svg}
\usepackage{rotating}
\usepackage{appendix}
\usepackage{graphicx}
\usepackage{natbib}
\usepackage{xspace}
\usepackage{color}
\usepackage{amsmath}
\usepackage[flushleft]{threeparttable}
\usepackage{hyperref}

\newcommand{\jyb}    {~Jy~beam$^{-1}$\xspace}
\newcommand{\mjyb}   {~mJy~beam$^{-1}$\xspace}
\newcommand{\kms}     {~km~s$^{-1}$\xspace}

\newcommand{\mjy}     {~mJy~beam$^{-1}$\xspace}

\newcommand{\msun}    {~$M_{\odot}$\xspace}
\newcommand{\lsun}    {~$L_{\odot}$\xspace}

\newcommand{\co}      {CO\,(2$-$1)\xspace}

\usepackage{lineno}
\nolinenumbers
\begin{document}

    \title{Confirming the Explosive Dispersal Outflow in DR21 with ALMA}

     \author{E. Guzm\'an Ccolque\inst{1}
          \and
          M. Fern\'andez L\'opez\inst{1}
          \and
          L. A. Zapata\inst{2}
          \and
          J. Bally\inst{3}
          \and
          P. R. Rivera-Ortiz\inst{2}
          }

    \institute{Instituto Argentino de Radioastronomía (CCT- La Plata, CONICET, CICPBA, UNLP), C.C. No. 5,1894, Villa Elisa, Buenos Aires, Argentina
        \and
        Instituto de Radioastronomía y Astrofísica. Universidad Nacional Autónoma de México, 58090, Morelia, Michoacán,México
        \and
        Center for Astrophysics and Space Astronomy, Department of Astrophysical and Planetary Sciences University of Colorado, Boulder, CO 80389, USA}

    \date{Received December XX, 2023; accepted  XX, 2024}    

    \abstract
    {We present Atacama Large Millimeter/submillimeter Array (ALMA) 1.3\,mm continuum and \co line emission observations toward the high-mass star formation region DR21. Five new continuum sources are found. We identify eighteen outflow streamers detected in CO emission radially arising from a common origin. The velocity spread of the outflow streamers range between $-100$ to $+70$\kms. The radial velocities of each outflow roughly follow linear gradients (Hubble–Lemaitre–like expansion motions). Using the CO emission of the whole ensemble of streamers we estimate a total outflow mass of 120--210\msun. Additionally, we derived the dynamical age (8600\,yr), momentum ($\sim10^{3}$\msun\kms), and kinetic energy ($\sim10^{48}$\,erg) of the outflow. The morphology and kinematics presented by the CO outflow streamers confirm the presence of an explosive dispersal outflow at the heart of DR21. Five dispersal explosive outflows associated with massive star-forming regions have been confirmed in our Galaxy (Orion BN/KL, G5.89-0.39, S106-IR, IRAS16076-5134 and IRAS 12326-6245). However, their frequency of occurrence in the Galaxy and the originating nature are still uncertain.}


    \maketitle

\section{Introduction} \label{sec:intro}
Explosive dispersal outflows have become a new phenomenon observed in the star-forming regions of massive stars \citep{2017_Bally,2020_Zapata,2022_Guzman}. They present morphological and kinematic features that differentiate them from classical bipolar outflows \citep{2017_Zapata}. The explosive dispersal outflows seem to be impulsive, and possibly created by an energetic single and brief event \citep{2009_Zapata}. These outflows consist of dozens of collimated CO outflow streamers \footnote{We will use the words streamers or filaments to refer to narrow straight molecular collimated ejections. Not to be confused with filamentary molecular clouds \citep{2014_Fernandez} or accretion streamers \citep{2023_Fernandez}.}, [FeII] fingertips, and H$_{2}$ wakes pointing back approximately to a central position \citep[see the case of Orion BN/KL,][]{2009_Zapata,2017_Bally,2023_McCaughrean}. The CO streamers are quasiradially distributed and appear nearly isotropic in the sky. They present well defined linear velocity gradients, resembling Hubble–Lemaitre velocity laws. In the case of Orion BN/KL, the dynamic ages of most CO streamers are close to 550 years, coincident with the age of the probable disruption of a non-hierarchical massive and young stellar system \citep{2005_Gomez,2008_Gomez,2020_Bally}. The location of the disruption of this stellar system could be traced down from the positions and proper motions of the main young stellar objects found to recede from each other \citep{2005_Rodriguez,2005_Gomez,2020_Bally}. The probable cause that lead to the ejection of these stars is a dynamic interaction between the members of the original compact stellar system \citep{2011_Bally}. 
The nature of dispersal explosive outflows is still under debate, however, a few ideas have been proposed, such as a protostellar merger or the disintegration of a disk in a close encounter \citep{Raga2021,Rivera2021}. Henceforth, the origin of the explosive event and its causes can only be revealed by meticulously analyzing the stellar remnants and the streamers, including their positions, velocities, masses, disks, jets, and other pertinent factors \citep{Rivera2019} that could be reproduced by numerical models \citep{Rodriguez2023}. ALMA observations showed several well-defined collimated molecular streamers in Orion BN/KL \citep{2017_Bally}, G5.89-0.39 \citep{2020_Zapata}, IRAS\,16076-5134 \citep{2022_Guzman} and, IRAS\,12326-6245 \citep{2023_Zapata}. \citet{2013_Zapata} reported another explosive dispersal outflow in DR21 and, more recently, \citet{2022_Bally} found an additional one in Sh106-IR. The center of the explosions in G5.89-0.39, DR21 and Sh106 are dominated by the ionizing emission of ultra-compact and more evolved HII regions, hiding the originating sources and hampering the interpretation of the explosion origin at millimeter wavelengths. The estimated rate of explosive dispersal outflows (based in the few cases identified) is comparable to that of the supernovae \citep[one event every $\sim$100 years,][]{2022_Guzman}, which suggest that dispersal outflows may have a large impact in several respects in the way some massive stars form or evolve during their early times.

DR21 is a region of massive star formation that is part of the Cygnus X region \citep{1993_Chandler}. According to measurements made with the Very Long Baseline Array using trigonometric parallaxes and proper motions of methanol and water masers, the distance to DR21 is 1.5$\pm$0.07\,kpc \citep{2012_Rygl}. \citet{1978_Dickel} proposed that, at large scales, the W75 cloud and the DR21 cloud are interacting. The target region of DR21 comprises a number of compact HII regions \citep{1973_Harris} and a well-studied classical outflow that is associated with them \citep[e.g.][]{1991a_Garden, 1992_Russell}. Two cometary HII regions are observed in DR21, one compact in the north (D) and the other, more extended, in the south, which has been resolved into three compact condensations \citep[A, B and C,][]{1973_Harris,1989_Roelfsema,2003_Cyganowski}. DR21 produces a powerful and very prominent in infrared molecular bipolar outflow \citep{1991b_Garden,1992_Garden,2007_Cruz}. It exhibits shocked molecular hydrogen emission which is extended in an east-west direction, with the red-shifted material being found to the east \citep{1991a_Garden,1996_Davis,1998_Smith}. The outflow shocks are thought to produce methanol abundance enhancements responsible for the detection of maser emission \citep{1997_Liechti}. Recent observations with NOEMA show the spatial distribution of several molecular tracers, including HCO$^{+}$, HCN, HNC, N$_{2}$H$^{+}$, H$_{2}$CO, and CCH. The HCO$^{+}$ emission shows blue- and red-shifted lobes overlapped \citep{2023_Skretas}. This outflow is among the most massive and energetic observed in the Galaxy \citep{1991b_Garden}. \citet{2013_Zapata} suggested the existence of an additional explosive outflow coming out from the same compact HII regions driving the classical bipolar outflow. This explosive outflow took place about 10,000 years ago. \co Submillimeter Array observations revealed about 25 molecular blue-shifted and red-shifted collimated streamers. All streamers appear to point toward a common center. The radial velocity along each streamer roughly follows Hubble-Lema\^{i}tre laws.\\

Using high angular resolution and sensitive Band 6 ALMA observations, we study the CO molecular emission of the DR21 outflows. We detect emission from the bipolar outflow and we confirm the explosive nature of nineteen outflow streamers. The paper is organized as follows. In Section \ref{sec:observations}, we present the ALMA observations. The identification of the 1.3\,mm continuum sources, the \co outflow streamers, and the new location of the origin of these streamers are presented in Section \ref{sec:results}. A brief discussion about the nature of the explosive dispersal outflow and the classical bipolar outflow in DR21 can be found in Section \ref{sec:discussion}. The conclusions of this study are presented in Section \ref{sec:conclusion}. 

\section{ALMA observations} \label{sec:observations}

The observations obtained with ALMA (Atacama Large Millimeter/submillimeter Array) consist of a mosaic with the 12m-antennas array in Band 6 on DR21. The mosaic was constructed with 41-pointing distributed in a Nyquist-sampled grid, and carried under program 2019.1.00263.S (P.I; John Bally). DR21 was observed in two sessions (see Table \ref{tab:observations2019}) during 29 April 2021 (with 42 antennas) and 6 May 2021 (with 44 antennas), totaling approximately 100 minutes on source. The maximum baselines for the two observing sessions were 1.3 and 2.5\,km, respectively. The average precipitable water vapor was 0.2\,mm and 1.1\,mm, and the average system temperatures 83\,K and 113\,K for both dates, respectively. The coordinates of the mosaic center were $(\alpha,\delta)_{J2000.0}=20^{h}39^{m}01.256^{s},+42^{\circ}19^{'}30.7^{''}$.

\begin{figure}
    \centering
    \includegraphics[scale=0.6]{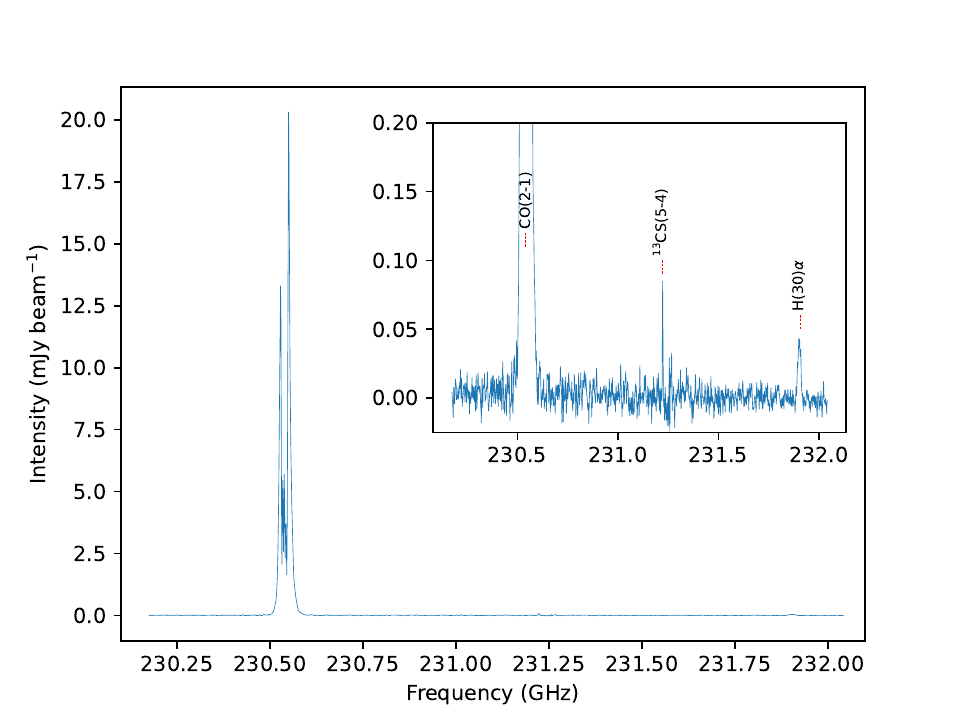}
    \caption{ALMA spectrum of the spectral window centered at 231.121\,GHz. The CO line at 230.54\,GHz can be clearly identified with a peak intensity of $\sim$20\mjyb. A deep absorption of the line caused by the cloud where DR21 is located is observed. In addition, two lines of lower intensity ($\sim$10 orders of magnitude less) are identified: $^{13}$CS\,(5$-$4) at 231.222\,GHz and H(30)$\alpha$ at 231.908\,GHz.} 
    \label{fig:spectrum}
\end{figure}

For all observations, the spectral windows were centered at 231.121, 217.704 and 220.012\,GHz so that they cover the transitions \co, SiO\,(5$-$4) and $^{13}$CO\,(2$-$1) at 231.121, 217.704 and 220.012\,GHz, respectively. The fourth spectral window was centered at 233.013\,GHz for continuum purposes. In this work, we focus on the 1.3\,mm continuum and the \co line emission (see Figure \ref{fig:spectrum}). J2007+4029 was used as the phase calibrator, J2253+1608 and J1924-2914 both as the bandpass and flux calibrator.

The data were calibrated following the standard procedures using the pipeline provided by the ALMA staff. The continuum image was constructed using the \texttt{tclean} task of CASA version 6.4.3.27, whereas the \co cube was directly downloaded from the ALMA Science Archive Products. The cleaning process of the continuum image was performed manually, by applying the H\"ogbom algorithm with a Briggs weighting, and by setting the robust parameter to 0.5. The rms noise level in the continuum image is 0.06\mjyb and the angular resolution is $0\farcs74\times0\farcs29$ (PA$=1.6\degr$). 
Regarding the CO velocity cube from the ALMA archive, its original angular resolution is 1.26\kms. We measured an rms noise level of 5\mjyb per channel, and the angular resolution is $0\farcs66\times0\farcs24$ (PA$=3.3\degr$). We further prepared a 5-channel binned version of the CO cube with 6.3\kms channels, which we use to estimate the outflow mass and energetics (Section \ref{sec:outflowid}).  

\begin{table*}
\small
\caption{Summary of Band 6 ALMA Observations of DR21 (Project 2019.1.00263.S)}
\label{tab:observations2019} 
\centering
\begin{tabular} {c|c|c}
\hline \hline
Execution Blocks IDs & uid://A002/Xeb9695/X4a88 & uid://A002/Xebc2ec/X2c12 \\
\hline
    Observation date & 2021-04-29 & 2021-05-06 \\
    Number of antennas & 42 & 44 \\
    Time on Source (sec) & 3002 & 2999 \\
    Number of pointings & 41 & 41 \\
    Mean PWV(mm) & 0.22 & 1.06 \\
    Phase Calibrator & J2007+4029 & J2007+4029 \\
    Bandpass Calibrator & J2253+1608 & J1924-2914 \\
    Flux Calibrator & J2253+1608 & J1924-2914 \\
\hline
\end{tabular}
\end{table*}

\section{RESULTS} \label{sec:results}

\subsection{Millimeter Continuum Sources}

\begin{figure*} [h!]
\centering
\includegraphics[scale=0.7]{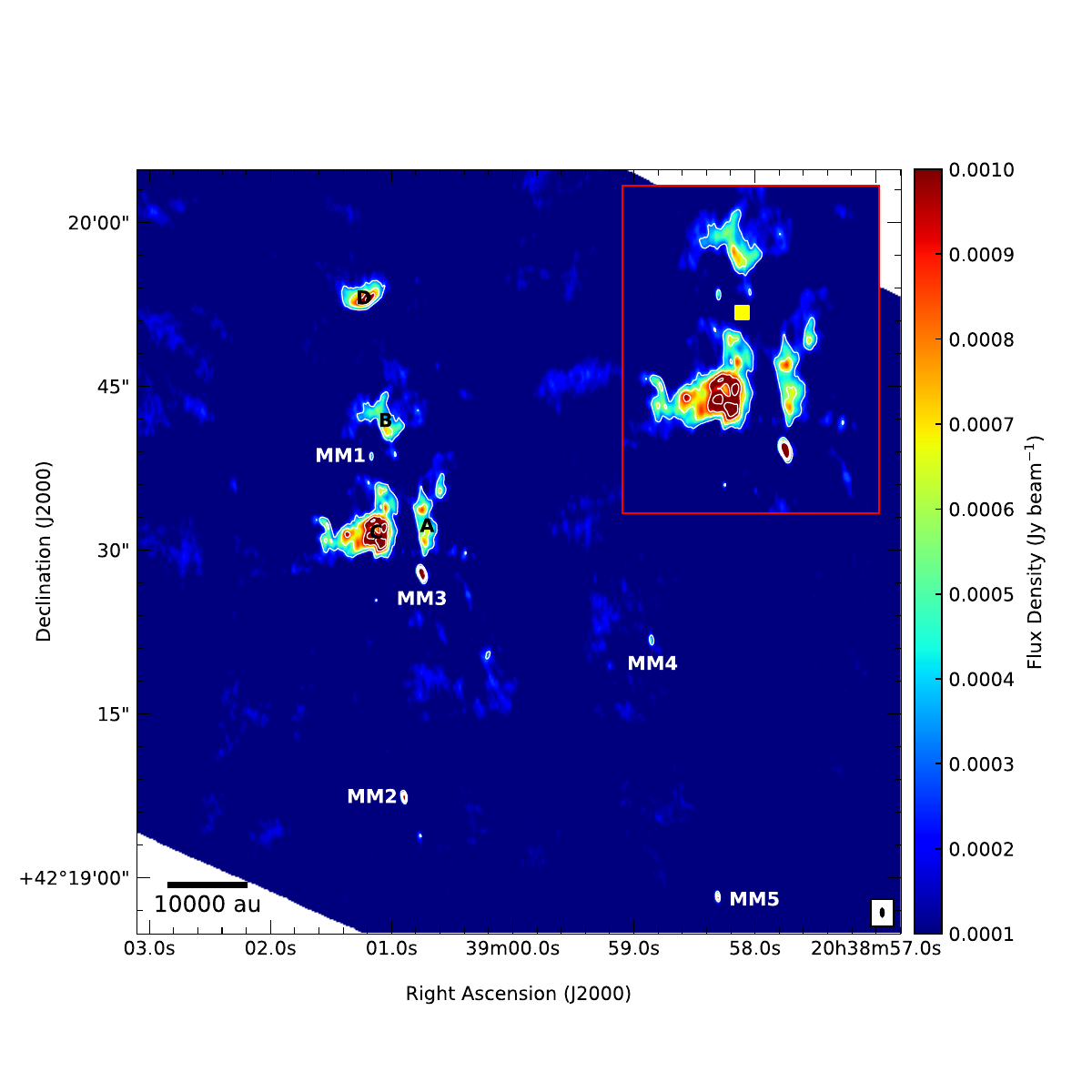}
\caption{1.3\,mm continuum emission toward DR21 in colour-scale and white contours (at 5, 10, 15, and 20 times the rms noise level of 0.06\mjyb). The synthesized beam, at the bottom right corner, is $0\farcs74\times0\farcs29$ (PA$=1.6\degr$). The new millimeter continuum compact sources reported here are labeled MM1-MM5, while the cometary sources follow the notation used in \citet{1973_Harris}. The central squared region is zoomed-in at the top right inset, which includes the position of the possible explosion center marked with a yellow square (see Section \ref{sec:outflowcenter}).}
\label{fig:cores}
\end{figure*}

Figure \ref{fig:cores} shows the positions of ten 1.3\,mm continuum sources detected with ALMA in DR21. We consider structures with sizes equal to or larger than the beam size and flux densities greater than 5$\sigma\sim$0.3\mjyb as criteria for identifying continuum sources (here $\sigma$ is estimated analyzing the empty background emission). The continuum map mainly shows two extended emission sources (of cometary appearance) and some compact sources. The extended sources correspond to the two well-known compact HII regions. The southernmost comprises fragments labeled as A, B, and C, and the northern is labeled D. All of these components have already been identified, and their emission at 1.3\,mm is most likely to be associated with free-free emission \citep{1973_Harris,2003_Cyganowski}. Five new compact sources, with effective radius $>$0.7\,\arcsec (beam's major axis), are identified in this work. This new sources are labeled with numbers sorted by their Right Ascension. Table \ref{tab:continuum} presents the position, source size, peak intensity, and integrated flux of these continuum compact sources. These parameters, and their uncertainties were estimated fitting 2D-Gaussians by using the CASA task \texttt{imfit}.

\begin{table*} 
\small
\caption{Submillimeter Continuum Sources in DR21}
\label{tab:continuum}
\centering
    \begin{tabular}{c|c|c|c|c|c}
    \hline\hline  
    Source & RA & DEC & Deconvolved Size & Peak Intensity & Flux integrated  \\
     & [ICRS] & [ICRS] & [$\arcsec\times\arcsec,\degr$] & [\mjyb ] & [mJy] \\
    \hline
    MM1 & 20:39:01.16573$\pm$0.00018 & +42:19:38.57532$\pm$0.01196 & point source & 0.55$\pm$0.01 & 0.59$\pm$0.02  \\
    MM2 & 20:39:00.89561$\pm$0.00031 & +42:19:07.36202$\pm$0.01371 & point source & 1.80$\pm$0.05 &  2.2$\pm$0.1  \\
    MM3 & 20:39:00.7460$\pm$0.0009 & +42:19:27.78$\pm$0.02 & 0.58$\pm$0.07$\times$0.17$\pm$0.07,38$\pm$8 & 3.6$\pm$0.1 & 6.4$\pm$0.4  \\
    MM4 & 20:38:58.8530$\pm$0.0006 & +42:19:21.72$\pm$0.02 & 0.7$\pm$0.2$\times$0.30$\pm$0.01,12$\pm$15 & 0.51$\pm$0.04 &  1.0$\pm$0.1 \\
    MM5 & 20:38:58.3048$\pm$0.0002 & +42:18:58.27$\pm$0.01 & point source & 0.95$\pm$0.02 &  1.04$\pm$0.05  \\
    A & 20:39:00.746$\pm$0.002 & +42:19:33.74$\pm$0.03 & extended source & 0.77$\pm$0.04 &  10.0$\pm$0.5  \\
    B & 20:39:01.034$\pm$0.003 & +42:19:41.04$\pm$0.04 & extended source & 0.74$\pm$0.05 &  13$\pm$1  \\
    C & 20:39:01.123$\pm$0.005 & +42:19:31.57$\pm$0.05 & extended source & 1.22$\pm$0.06 &  90$\pm$5  \\
    D & 20:39:01.208$\pm$0.007 & +42:19:53.00$\pm$0.04 & extended source & 1.12$\pm$0.05 &  24$\pm$1  \\
    \hline
    \hline
    \end{tabular}
    \begin{tablenotes}
        \small {
        \item{Note: The parameters of submillimeter continuun sources, and their uncertainties were estimated fitting 2D-Gaussians by using the CASA task \texttt{imfit}.}}
    \end{tablenotes}
\end{table*}

\subsection{Outflow Identification, Kinematics and Energetics} \label{sec:outflowid}
\begin{figure*}
\centering
\includegraphics[scale=0.6]{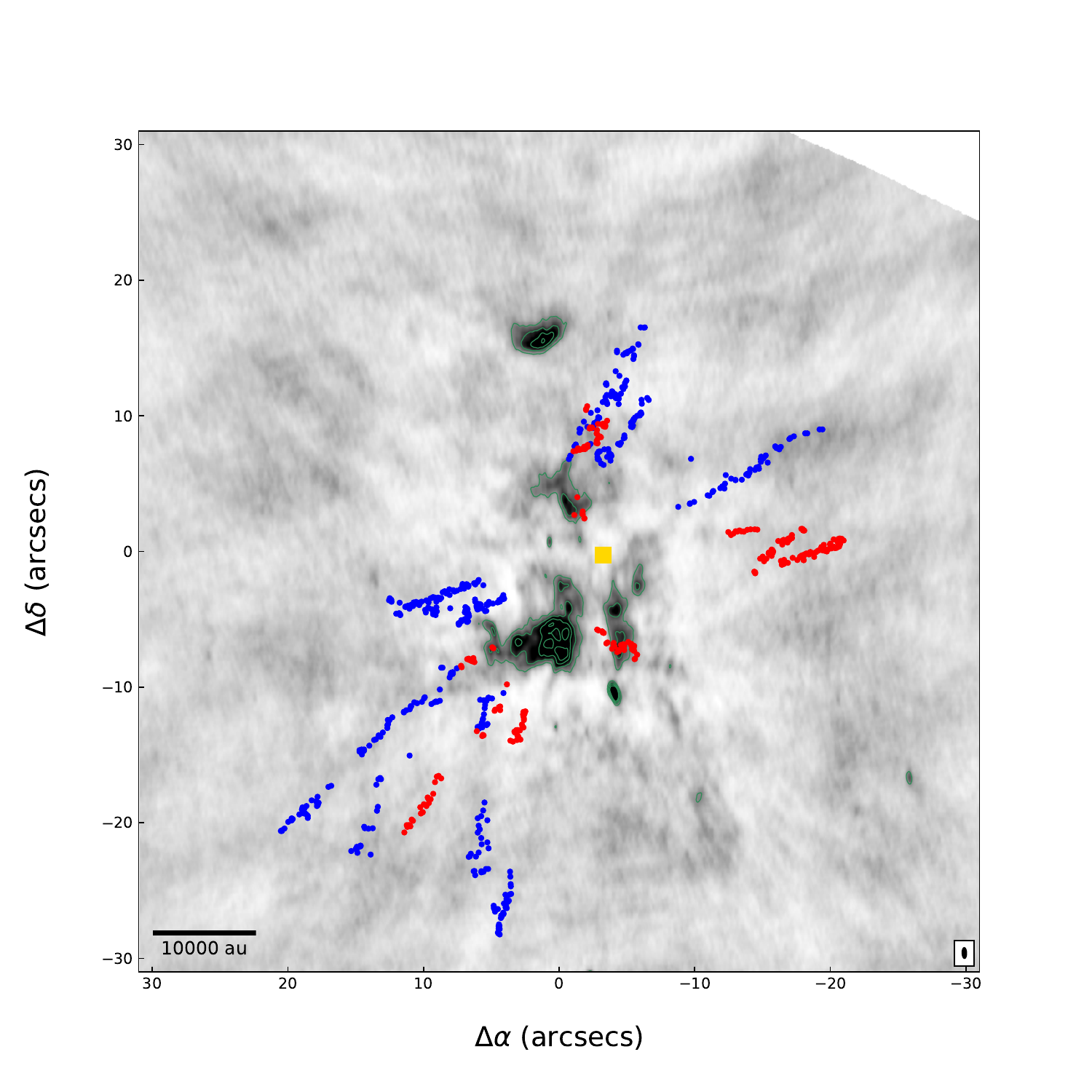}
\caption{The image shows the 1.3\,mm continuum emission with an overlay including the position of \co condensations in DR21. The background gray scale image and the green contours (at 5, 10, 15, and 20 times the rms noise level of 0.06\mjyb) are the ALMA 1.3\,mm continuum image from Figure \ref{fig:cores}. The red and blue circles correspond to position of red-shifted and blue-shifted \co gas condensation, identified in the velocity cube. The yellow square marks the position of the explosion center. The synthesized beam is shown within a box at the bottom right corner. The central position is at (0\arcsec,0\arcsec), where in RA it is at $20^{h}39^{m}01^{s}.1$ and DEC it is $+42^{\circ}19^{'}37^{''}.9$.}

\label{fig:condensations+cont}
\end{figure*}

\begin{figure}
    \centering
    \includegraphics[scale=0.65]{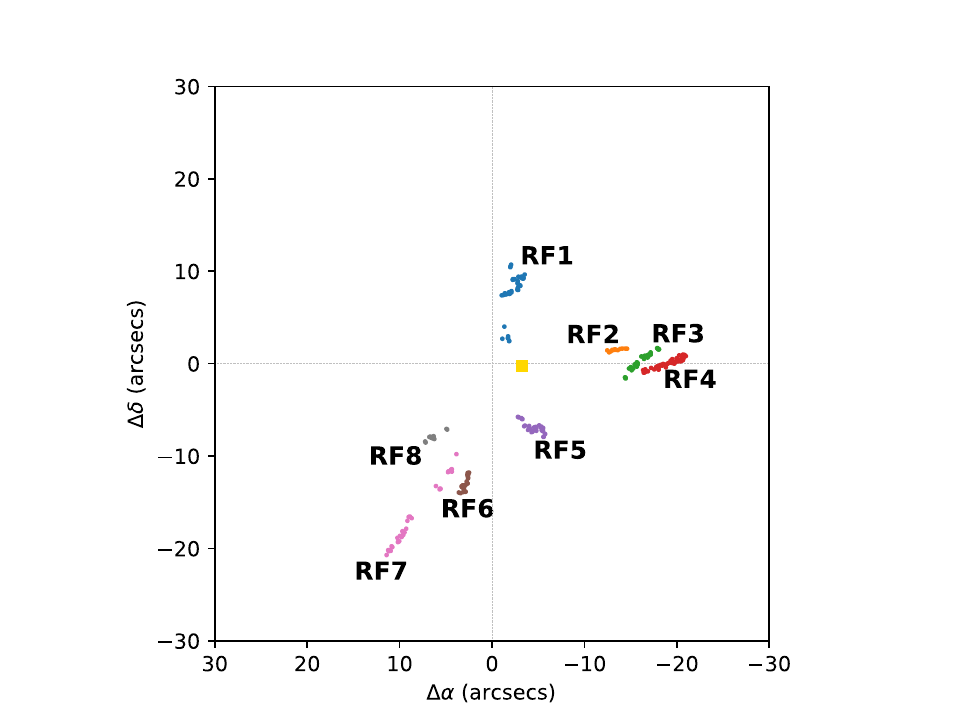}
    \hspace{1mm}
    \includegraphics[scale=0.65]{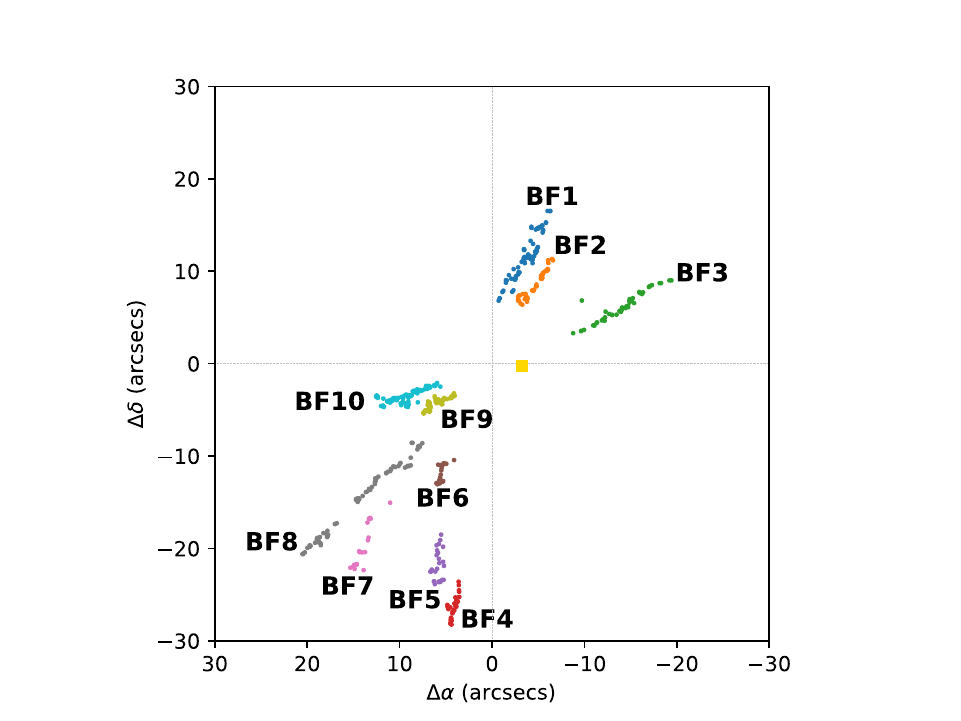}
    \caption{Red-shifted (left panel) and blue-shifted (right panel) \co gas condensations detected in DR21. The blue-shifted emission has radial velocities from $-$25 to $-$101\kms, while red-shifted emission exhibits radial velocities from +20 to +69\kms. The velocities are expressed with respect to the molecular cloud velocity. We identified 9 red-shifted filaments and 10 blue-shifted, labeled RF and BF, respectively. The possible origin of all the CO streamers is marked with the yellow square. }
    \label{fig:blue-red}
\end{figure}

\begin{figure*}
    \centering
    \includegraphics[scale=0.6]{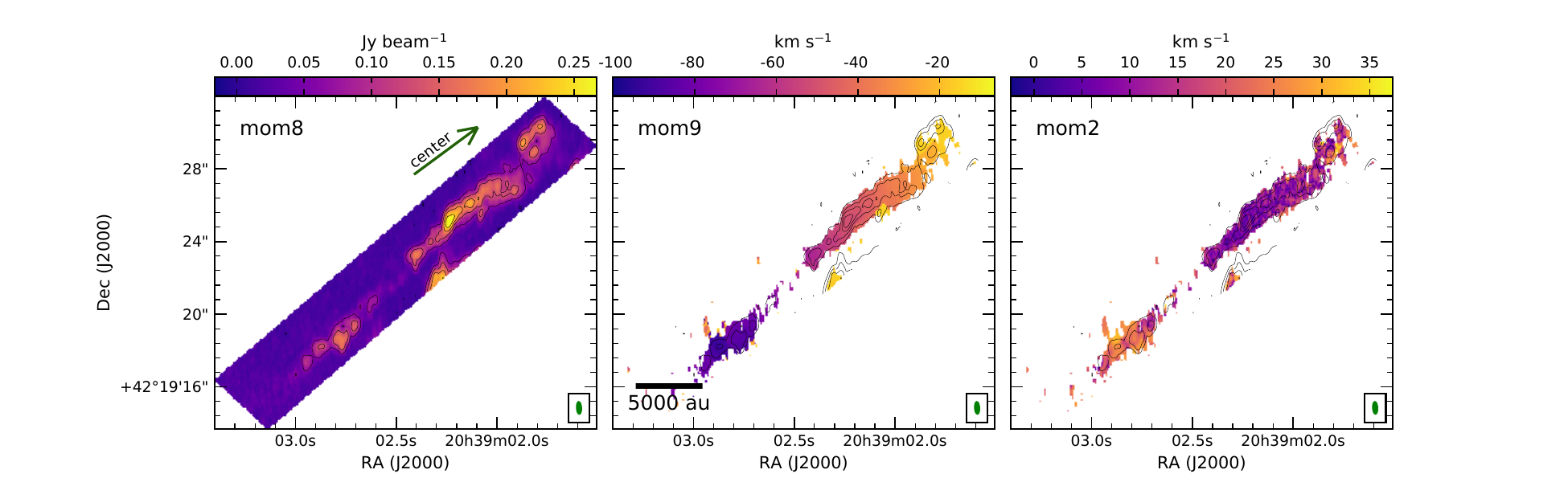}
    \caption{The moment-8, moment-9 and moment-2 maps of \co line emission for the streamer BF8 (left, medium and right, respectively). The black contours represents the \co emission at 3, 5, 8 and 10 times the rms noise level of 0.2\jyb in the moment-8 map. The synthesized beam is shown within a box at the botton right corner.}
    \label{fig:BF8_moms}
\end{figure*}

\begin{figure*}
    \centering
    \includegraphics[scale=0.7]{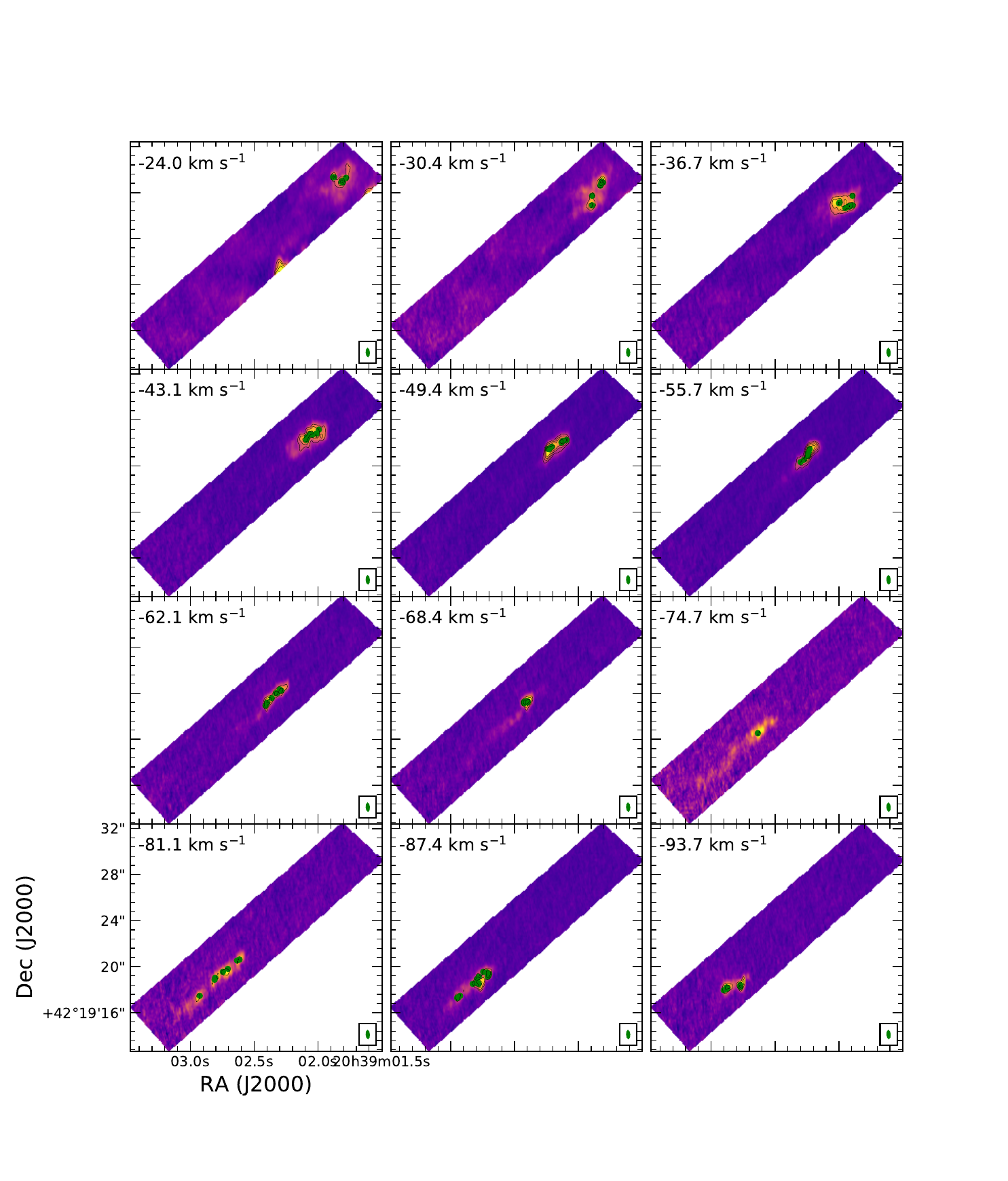}
    \caption{co emission velocity channels of the streamer BF8. The original channels were binned to produce a 5-channel binned cube with binned channels of 6.34\kms in width. The black contours represents the emission at 4, 6, 8, and 10 times the rms noise level of 20\mjyb. The green dots are the condensations identified in the streamer BF8 (see Section \ref{sec:outflowid}). }
    \label{fig:BF8_panel}
\end{figure*}

Our high angular resolution \co observations revealed new molecular filaments or streamers with different orientations in the inner region of DR21 (see Figures \ref{fig:condensations+cont}, \ref{fig:blue-red}, \ref{fig:BF8_moms}, \ref{fig:BF8_panel}, and also Appendix \ref{appx:cubos}). Some of these streamers agree very well with those found by \citet{2013_Zapata} using the SMA, analyzing a field of view three times larger than the data presented here taken with ALMA. All of these molecular streamers appear to emerge from a common center (yellow square in Figures \ref{fig:condensations+cont} and \ref{fig:blue-red}; also see Section \ref{sec:discussion}) and are contained in a circle of radius $\sim$30$\arcsec$ ($\sim$0.2\,pc, taking the distance to DR21 to be 1.5\,kpc). The position of the center does not coincide with any of the continuum sources. In Figure \ref{fig:condensations+cont}, we show, overlapped on top of the 1.3\,mm continuum emission (grey scale and green contours), the \co molecular emission detected in the ALMA mosaic toward DR21 as colored red/blue circles. As criteria for identifying the circles, hereafter called condensations, we use structures with CO emission above 20\mjy (emission over 4$\sigma$ threshold), and sizes similar to beam's minor axis. Condensations were identified after careful manual inspection of the velocity cube on a channel-by-channel basis. We avoided the cloud velocity channels (between $-26$ and +21kms), which were heavily contaminated by the presence of strong extended emission and also presented artefacts and blanked channels, caused by interferometric missing flux. The position and the radial velocity of each condensations, were obtained using the Miriad's task \texttt{cgcurs} \citep{1995_Sault} and added to a list of CO condensations. The \co blue-shifted emission ranges in radial velocity from $-$25 down to $-$101\kms, whereas the red-shifted emission goes from $+20$ up to $+69$\kms. After this process of identifying condensations through all the CO cube, we were able to visually identify some conspicuous linear chains of condensations. These chain are coherent both spatially and in radial velocity. We call these filamentary-like structures, streamers. In Figure \ref{fig:blue-red} we label 8 red-shifted (RF1,..., RF8) and 10 blue-shifted streamers (BF1,...,BF10). 
Some of the streamers, both red-shifted and blue-shifted, are intertwined in the plane of the sky.
They appear distributed more or less radially with respect to a common center, and, assuming this center as their origin, none of them seem to have a bipolar counterpart. It is worth to note that there is little or almost no CO emission in the NE-SW direction (see Section \ref{bipolar}). Also, except for the low-velocity, central channels of the cube, no \co condensations are observed toward the central part of the map. This is because the central channels of the velocity cube, where the parent cloud emission predominates, have been omitted.

The moment-8, moment-9 and moment-2 maps \footnote{In general, the word \lq moment\rq\, refers to collapsing an axis of a 3D dataset in a definite way to form a 2D image. For example, given a RA-DEC-Velocity cube, collapse the velocity axis by computing the mean intensity of the spectrum at each RA-DEC pixel. The moment-2 map shows the intensity weighted dispersion of the radial velocity and it is traditionally used to get "velocity dispersion", the moment-8 map shows the maximum intensity value of the spectrum (the peak), and the moment-9 map, the corresponding radial velocity at the peak value of the spectrum. See section 1.1.1 in the CASA Toolkit Reference Manual. In this work, we prepared the moment maps using the CASA task \texttt{immoments}.} of the \co line emission for the BF8 streamer are shown in Figure \ref{fig:BF8_moms}, from left to right. We choose these moment maps over the more classical moment-0 (integrated intensity) and moment-1 (intensity weighted velocity), because they better stress the filament morphology and velocity gradients of these structures. In particular, the BF8 filament morphology is one of the the most clear examples in DR21, and we use it to exemplify the main properties of these structures.
BF8 runs south-east for about $16\arcsec$ (24000\,au at the assumed distance to DR21), following a straight path, and comprises more than 50 condensations. Its width varies between $0\farcs75$ to $2\farcs25$ (that is, 1000-3500\,au). Its moment-9 map (Figure \ref{fig:BF8_moms}) shows a trend of increase in velocity from the center (yellow square in Figure \ref{fig:condensations+cont} ; see also the yellow arrow in the moment-8 image) to along of streamer. The moment-2 map describes the dispersion of the CO line, revealing a very wide linewidth ($\sim 30$\kms) at the tip end of the filament, possibly linked to a zone with a strong shock. Figure \ref{fig:BF8_panel} shows some channels of the \co velocity cube of the region where we identified BF8. This velocity cube (between $-31.8$ and $-101.1$\kms) was constructed by applying a 5-channel binning in the spectral axis of the original cube. Hence, the final spectral resolution became 6.3\kms. A visual inspection of the velocity cube evidences the existence of a clear velocity gradient, with CO condensations moving away (to south-east) from the center proposed with increasing the velocity. For each of the red-shifted and blue-shifted streamers identified (see Fig. \ref{fig:blue-red}), we made figures analogous to Figure \ref{fig:BF8_moms} that can be found in the Appendix \ref{appx:cubos}.

\begin{figure*}
  \centering
  \includegraphics[scale=1.25]{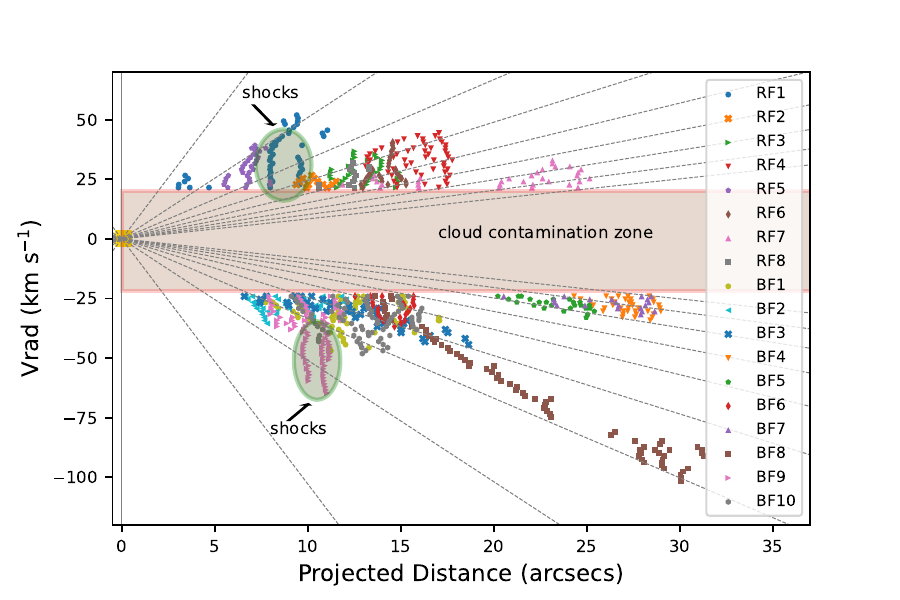}
  \caption{PV diagram of the \co emission of the set of filaments associated with DR21. Each filament is identified by a different colour and a different marker. The zero point in the diagram (yellow square) corresponds to the new centre measured in this paper. The grey lines represent linear trends between projected distance and radial velocity. All grey lines start at a projected distance of 0\arcsec and a radial velocity of 0\kms. They were constructed by increasing the slope every 5$^{\circ}$, in two ranges from 40$^{\circ}$ to 90$^{\circ}$ and -40$^{\circ}$ to -90$^{\circ}$. Some streamers appear to follow linear trends, while others show areas of large velocity dispersion. These can be explained as shock zones (green ellipses) between the streamer and its surrounding material. The pink rectangle indicates the region where it was not possible to distinguish the streamer emission due to contamination of the parent cloud.}
  \label{fig:gradiente}
\end{figure*}

We present a position-velocity diagram of all DR21 filaments in Figure \ref{fig:gradiente}, plotting the projected distance from the center and radial velocity (expressed relative to the parental cloud velocity) of each condensation identified in Figure \ref{fig:condensations+cont}. Each filament is distinguished with a different color and marker. The origin of the diagram (yellow square) corresponds to the new center measured in this paper (see Section \ref{sec:outflowcenter} bellow). The gray lines represent linear trends between projected distance and radial velocity. In general, the streamers seem to  qualitatively follow these trends, indicating that the velocity of the condensations increase with distance from the common center. It is worth noting that this kinematic behaviour (a linear increase in velocity with distance from the center) is one of the most distinct signatures of explosive outflows \citep{2017_Zapata}. For some other filaments, however, the trend is not that clear, which maybe caused by condensations not describing the crest of the filamentary structures, by the presence of wide arches due to shocks, the CO tracing the sides/walls of the streamers instead of their crest, or because the overlapping of patches of two or more filaments, for also to the interaction with the surrounding material. Moreover, within a radius of $5\arcsec$ (75000\,au) from the common origin, we do not detect significant compact CO emission, except in the velocity range close to the cloud velocity. In that specific velocity range (i.e., from $-26$ to $+21$\kms) the cloud contamination hampers the identification of filaments. On the other end, that is at the highest velocities, the projection effects may not be favorable for a clear detection of the filament structures (e.g., if the structures point directly toward the observer, as expected in an isotropic ejection of finger-like outflows).

In view of the morphological and kinematic resemblance of the streamers identified in this paper in DR21 with those in other explosive outflows (such as those in e.g., Orion\,BN-KL outflow), we can confirm that the ensemble of eighteen streamers most likely constitutes part of a single outflow entity with a common origin and which we can assume was ejected instantaneously. Considering local thermodynamic equilibrium and an optically thin CO emission, we estimate the mass, momentum and energy of the explosive outflow. We calculate the total mass of the outflow as the sum of the masses of all the filaments in each channel of the CO velocity cube (see Appendix \ref{appx:masses}). Again, we avoid the central channels with cloud contamination and/or missing extended emission.  The derived total mass ranges between 120 and 210\msun for 70\,K and 140\,K, respectively. We use this range of temperatures following the advice of F. Navarete \citep[priv. comm., and][]{2018Navarete,2019Navarete} based on observations of the CO~(7-6) and CO~(6-5) lines in a sample of 100 massive clumps. The derived temperatures can be used for outflows from the most luminous protostellar regions such as DR21 \citep[$10^5$--$10^6$\lsun][]{1991a_Garden,1992_Garden}. And using the same procedures for obtaining the total mass, we calculate the momentum ($\sim10^{3}$\msun\kms), and kinetic energy ($\sim10^{48}$\,erg) of the whole outflow, with values correct within a factor of 2-3. This implies that the outflow emission in DR21 is associated with a very energetic event \citep{2004_Wu}. Furthermore, taking $\sim$120$\arcsec$ as the most distant CO emission, that of the tip end of the DF1 filament reported by \citet{2013_Zapata}, we estimate a dynamical age of $\sim$8600 years for the outflow. This value is in agreement with the kinematic age presented in \citet{2013_Zapata}.

\subsection{Estimation of the center of the explosive outflow} \label{sec:outflowcenter}

\begin{figure*}
    \centering
    \includegraphics[scale=0.6]{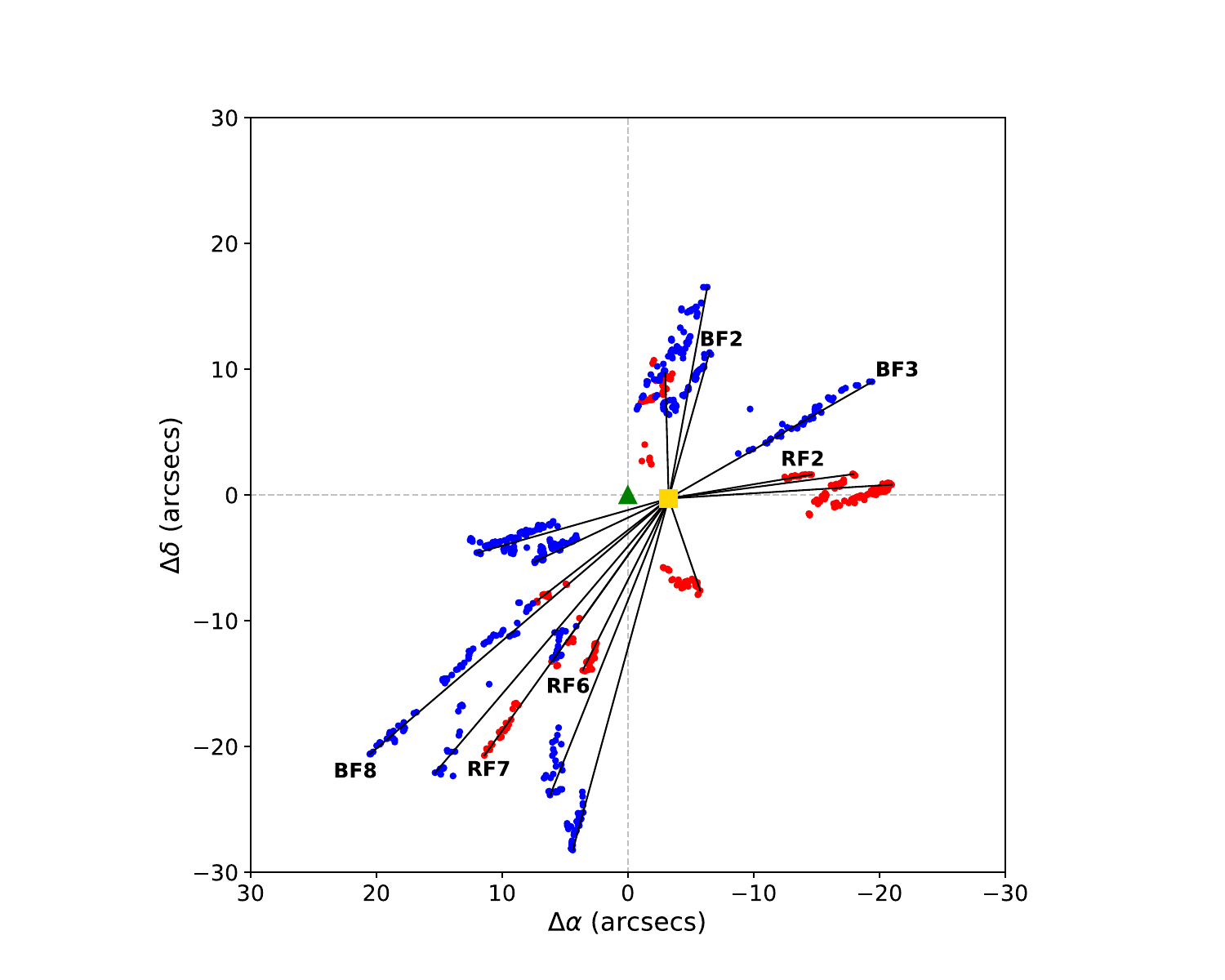}
    \caption{Image of the ensemble of CO streamer condensations identified in DR21. The red circles represent red-shifted emission while the blue circles represent blue-shifted emission. The green triangle shows the location of the center as determined by \citet{2013_Zapata}, whereas the yellow square reflects the new position described in this paper. The new position of the center is $20^{h}39^{m}00^{s}.8\pm00^{s}.2$ in RA and $+42^{\circ}19^{'}37.62^{''}\pm00^{''}.97$ in Dec. The  lines were constructed using linear fit for each streamer.}
    \label{fig:center}
\end{figure*}

Figure \ref{fig:center} shows the center of the explosive outflow (green triangle) found by \citet{2013_Zapata} and the new center (yellow square) found in this work. Joining the tip ends of the 18 CO streamers with the position of their common center, the drawn lines trace quite well the orientation and the path of the condensations of every streamer. 
 
\begin{figure}
    \centering
    \includegraphics[scale=0.55]{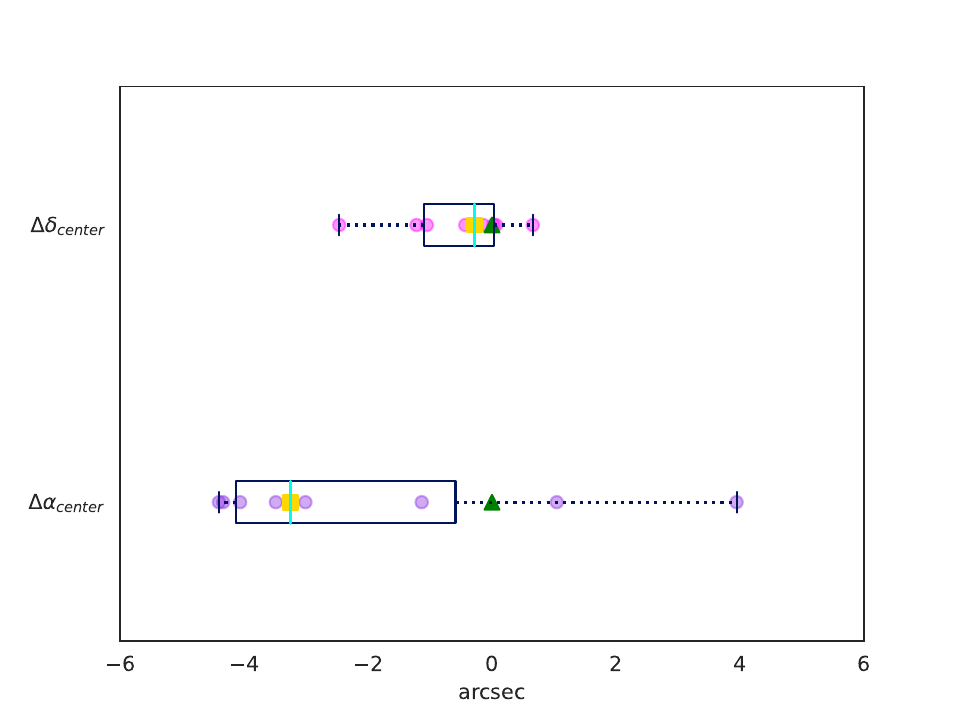}
    \caption{Statistical analysis of the center of the explosive outflow in DR21. Pink circles show the offsets in RA and Dec, with respect to the center reported by \citet{2013_Zapata}, of the intersections between pairs of different outflow trajectories (extracted as explained in the text). The position of the origin (0\arcsec, 0\arcsec) is set at $20^{h}39^{m}01.1^{s}$ in RA and $+42^{\circ}19^{'}37.9^{''}$ in Dec. In this paper, the median (yellow square) is considered as the new position of the center of the outflow. The vertical colored line represents the median value, the box represents the interquartile range between the 25th and 75th percentiles, and the vertical lines (the `whiskers') show the full range of the data without outliers.}
    \label{fig:boxplot}    
\end{figure}

To find out the position for the new center we used the 6 streamers with the most clear, straight, and well-defined trajectories: RF2, RF6, RF7, BF2, BF3 and BF8. For each of the 6 streamers a line was constructed by linear fitting of all their identified CO condensations (see Section \ref{sec:outflowid}). Then, we created a pool of intersection points with all the different pairs of streamers. We excluded those intersections that were more than 5\arcsec away from the center estimated by \cite{2013_Zapata}.
With all of this, the center of the explosive outflow was derived by computing the median of the intersection points (see Figure \ref{fig:boxplot}). The new center is set at $(\alpha,\delta)_{J2000.0}=(20^{h}39^{m}00^{s}.8\pm00^{s}.2,+42^{\circ}19^{'}37^{''}.62\pm0{''}.97$). This position lies within the expanding cometary HII region that contains the continuum sources A, B and C, just $\sim1\arcsec$ north of the edge of source C. \citet{2013_Zapata} suggested that both, the explosive flow and the large cometary HII region in DR21 comprising A, B, and C, were probably produced by the same mechanism. An analogous scenario is observed in the explosive outflow in G5.89-0.39 \citep[and also very possibly in IRAS\,12326-6245][]{2023_Zapata}, where there is an expanding shell centered at the origin of a similar ensemble of explosive outflow streamers \citep{2020_Zapata,2021_Fernandez-Lopez}. Both centers, the one estimated by \cite{2013_Zapata}, and the new one coincide very well within the errors.

\section{DISCUSSION} \label{sec:discussion}
\subsection{Confirmation of DR21 as an explosive dispersal outflow} \label{}
The high angular resolution ($\sim$0.7\arcsec) ALMA \co line emission observations toward DR21 show 18 collimated, high-velocity outflow streamers (with velocities in the range from $-$100 to $+$70\kms). These streamers appear to emerge quasi-radially from a common center, which is probably embedded within a cometary HII region.
Red-shifted and blue-shifted streamers are observed to be spatially overlapping in the plane of the sky. Moreover, there seems to be a linear trend between the radial velocity along the streamers and their projected distance in the plane of the sky. A kinematic energy of $10^{48}$\,erg is estimated for the whole outflow. This is in the range of the energies derived for previously observed explosive outflow events. Furthermore, there are no evidence of infrared sources at the location of the origin of these filaments, and we have not found any millimeter cores coincident with this position. If the DR21 event was similar to the Orion OMC1 event, one might expect that the stars involved were dynamically ejected and are now high-velocity runaway stars. All of these features are in agreement with those found as characteristic signatures in other explosive outflows, therefore confirming that the outflow found in DR21 is an explosive dispersal outflow, as suggested by \citet{2013_Zapata} from coarser CO(2$-$1) line observations made with the SMA.

\subsection{Explosive's outflow dynamical time and rate of events in the Milky Way} \label{sec:rate}
Until now, DR21 was thought to be the oldest explosive dispersal outflow found in the Milky Way. \citet{2013_Zapata} estimated its age in $\sim$10,000\,years. They assumed that the debris from the explosive event would have velocities of about 40\kms and
the that streamers are localized at an average distance of $80\arcsec$ ($1.2\times10^{5}$\,au). In this work, we estimate a similar dynamical time in a slightly different way. We consider all the ejected streamers move essentially at the same velocity, which translates into a range of maximum radial velocities for differently inclined streamers. We take 100\kms as the outflow, derived from the maximum velocity at which CO emission is observed. We assume this would be produced in a streamer ejected in a direction closer to the line-of-sight. However, this should be taken as a lower limit of the true velocity of the ejection.
For this estimate we assume all streamers move at the same constant speed since their ejection. Assuming the plasmon hypothesis from \cite{Rivera2019}, the material at the tip of the streamer decelerates as some mass is detached; however, if the density contrast between the ambient and the bullet is large enough ($\rho_{ambient}/\rho_{bullet}\leq10^{-2}$), and/or the mass detached from the plasmon is not reduced in a large percentage, the plasmon preserves almost the same velocity since the ejection. Instead, the upstream wake comprises detached and dragged material progressively decelerating, which explains the linear velocity gradients observed. For the outflow size, we use 120$\arcsec$, the value obtained by \citet{2017_Zapata}. In this way, we obtain an outflow dynamical age of $\sim8600$ yr. Note, that this should be treated as an upper limit, because of the velocity derivation aforementioned. Hence, following the same procedure as in the case of the IRAS\,16076-5134 outflow \citep[see][]{2022_Guzman}, the rate of explosive outflows in the Milky Way is still similar. It should be noted, however, that this is a very crude estimate in which the main uncertainties are the size of the Galaxy, the distance determination to the explosive outflows and the rough derivation of the outflow dynamical times. Therefore, our conclusion should be that the rate of explosive is of the same order of both, the supernova rate and the rate of formation of massive stars in the Milky Way.  

\subsection{About the main bipolar outflow in the region} \label{bipolar}
\begin{figure*}
    \centering
    \includegraphics[scale=0.59]{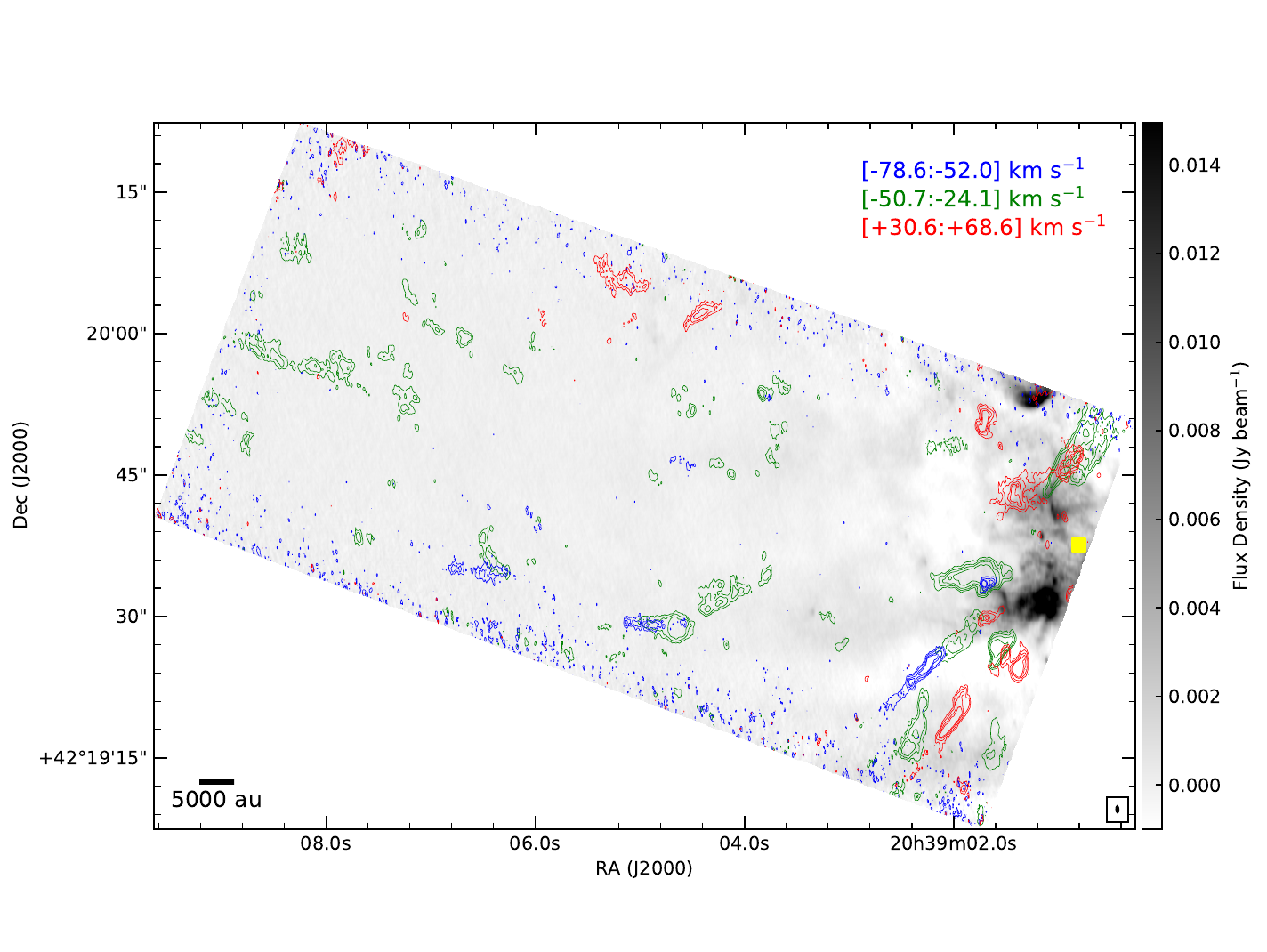}
    \hspace{1mm}
    \includegraphics[scale=0.59]{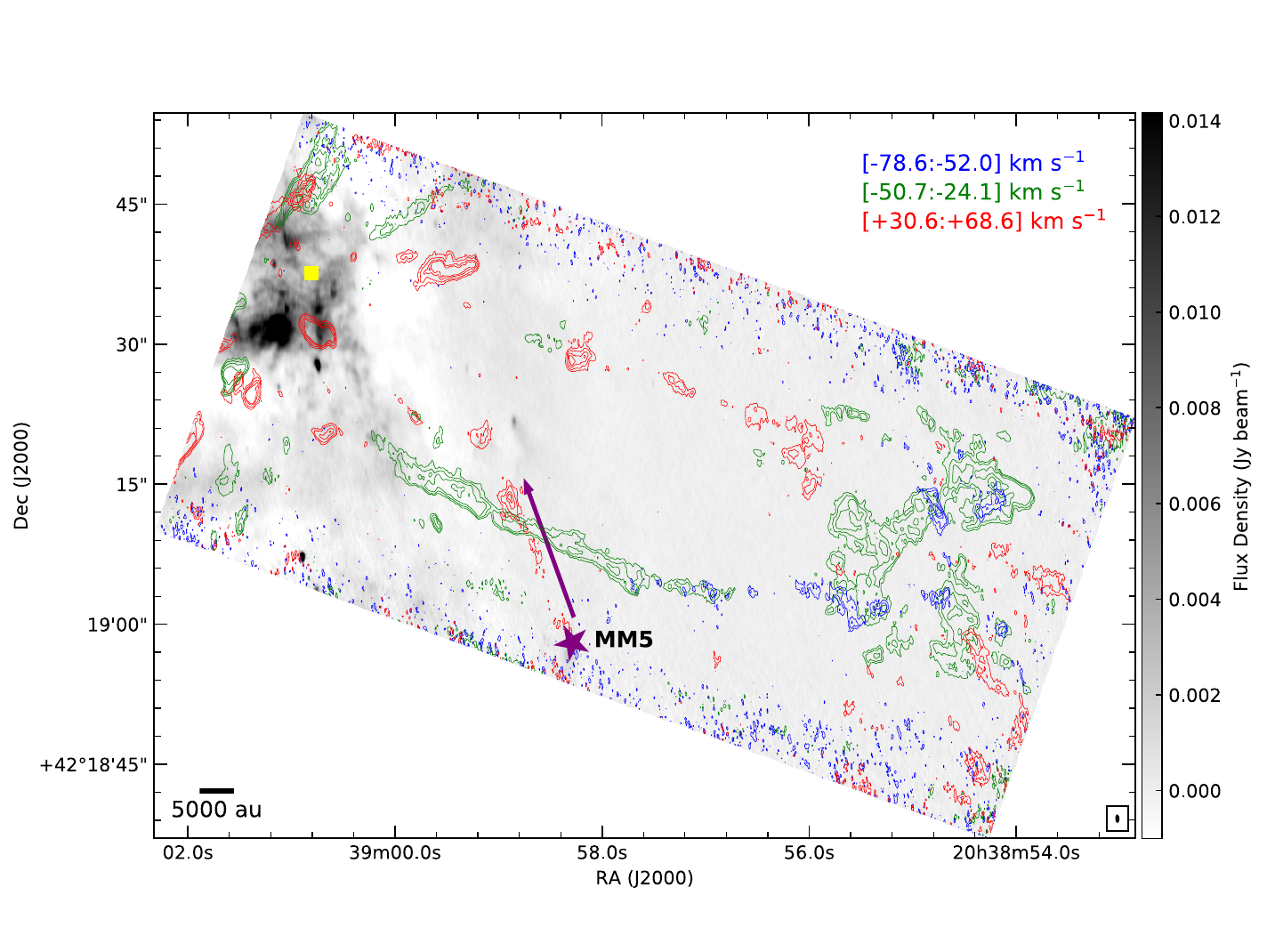}
    \caption{\co emission (contours) to the north-east (upper panel) and south-west (bottom panel) of the explosive outflow at DR21 overlaid with the continuum map at 1.3\,mm (grey scale). The coloured contours indicate different velocity ranges: blue for velocities between $-$79 and $-$52\kms, green between $-$51 and $-$24\kms and red between $-$31 and +69\kms. The contours represents the \co emission at 5, 10, 20,and 30 times the rms noise level of 70\mjyb \kms. A red-shifted collimated outflow is observed towards the southern edge of the image. The direction of this outflow (violet arrow) seems to indicate that it is associated with the continuum source MM5 (violet star). The synthesized beam of the continuum map is shown within a box at the bottom right corner.}
    \label{fig:bipolar_contornos_2}
\end{figure*}

Figure \ref{fig:bipolar_contornos_2} shows the \co emission detected with ALMA from the north-east/south-west classical bipolar outflow (contours) in DR21. Coloured contours overlay the 1.3mm continuum emission map (grey scale) showing different structures at different velocities: blue, for high velocity blue-shifted emission ([$-$79:$-$52]\kms ); green, for low velocity blue-shifted emission ([$-$51:$-$24]\kms); red, for red-shifted emission ([+31:+69]\kms). In comparison with the HCO$^+$ observations of \citet{2023_Skretas}, the ALMA observations do not show too much extended emission, but just ribbons and patches here and there. Actually, the north-east lobe is hardly recovered and we do not comment on it here. However, despite the preference for compact- and narrow-only features, the ALMA's images show some distinct structures toward the south-west lobe that can help us in shedding some light to interpret the nature of this classical bipolar outflow.

First, there are two clear quasi-parallel narrow $45\arcsec$ structures encompassing the body of the outflow. The southernmost edge is traced by green contours (mild blue-shifted velocities), while the northern one is less delineated and traced by red contours.  The two quasi-parabolic edges of the outflow close at a more chaotic tip \citep[called the interaction region in ][]{2023_Skretas}, about one arcminute from the center of the outflow.  This region shows a mixture of blue, green and red contours \citep[see also][reporting methanol maser emission at this place too]{1990_Plambeck}. \citet{2023_Skretas}, based on their molecular and kinematical analysis, discussed the possibility that the interaction region, with a mix of high and low red-shifted and blue-shifted velocities could be explained by the side-way motions of the gas produced in strong shocks from an outflow lying close to the plane of the sky.
  
We propose here an alternative scenario, in which there are several outflowing ejections beating up the so-called interaction region. As we will see, these ejections may be originated by the same protostar plus disk system, as it occurs in sources such Cepheus~A~HW2 or IRAS~15398-3359 \citep{2009_Cunningham,2021_Vazzano}. Molecular outflows comprising multiple outflowing ejections in different directions from the same protostar have been detected in massive stars such as Cepheus\,A\,HW2 and G5.90-0.39 \citep{2009_Cunningham,2013Zapata,2021_Fernandez-Lopez} and low-mass protostars such as IRAS\,15398-3359 \citep{2021_Vazzano,2024GuzmanCcolque}. A multiple system with close passages in highly eccentric orbits, or episodes of strong asymmetric accretion toward part of a disk have been proposed as the responsible for the periodic tilt of the ejection axis in these systems, producing a collection of bipolar outflowing ejections.

\begin{figure*}
    \centering
    \includegraphics[scale=0.6]{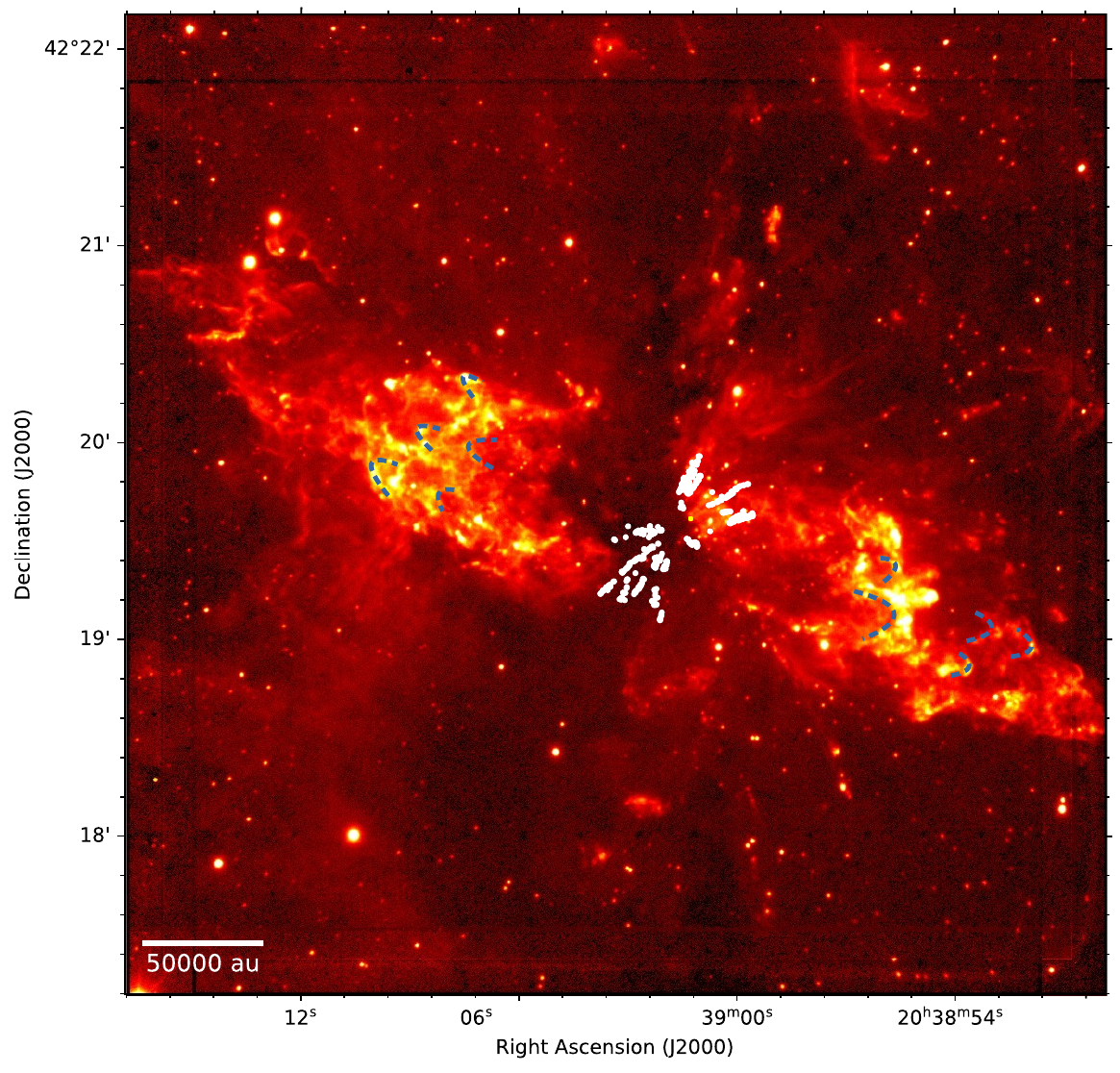}
\caption{2.12$\mu$m molecular hydrogen emission at DR21 overlaid with \co emission (white dots) of the explosive outflow. The H$_{2}$ emission was obtained using the Apache Point Observatory 3.5 meter telescope and NICFPS near infrared camera using a 0.5 bandpass filter (P.I., John Bally). The yellow square is the center of the outflow. The blue dash lines trace arc-shaped structures detected in H$_{2}$.} 
    \label{fig:bipolar-H2}
\end{figure*}

This scenario may reconcile several facts: (i) the very different radial velocities from the  cavity walls (a difference of at least 40\kms following Figure \ref{fig:bipolar_contornos_2}) may be explained if the outflowing ejections from the same source have slightly different inclinations with respect to the plane of the sky; (ii) the extremely wide outflow opening angle measured by \cite{2023_Skretas} would be, in this way, due to the action of several outflowing ejections with different position angles; (iii) the mixture of radial velocities in the interaction zone (Figure \ref{fig:bipolar_contornos_2}), with differences of more than 50\kms could be better explained if various outflowing ejections with different orientations are hitting the dense pocket of gas seen previously in N$_{2}$H$^+$ \citep{2023_Skretas}; (iv) last, but not least, the classical infrared H$_2$ images of the DR21 outflow from e.g. \citet{2007_Cruz}, and also the APO\,3.5\,m image (see Figure \ref{fig:bipolar-H2}
), show that there is a chaotic complex of arches and bow-shocks pointing in slightly different directions. These different arches can also be seen hinted in some of the blue-shifted channels of the HCO$^+$ cube presented in \citet{2023_Skretas}.

\subsection{Coexistence of the main bipolar outflow and the explosive outflow} \label{disgresion}
The ALMA data show that toward the DR21 region there are, on one hand, at least 18 outflow streamers with a common origin probably driven by an explosive event (the so-called explosive outflow). On the other hand, there are probably various bipolar outflowing ejections launched with slightly different orientations forming the main north-west/south-east bipolar outflow (the so-called bipolar outflow). Both outflows --explosive and bipolar-- are driven apparently from the same region, where the cometary HII regions lie.

The bipolar outflowing ejections have larger extent than the explosive streamers detected by ALMA. Also the ejections are not that well collimated and do not show linear velocity gradients. Although they show very fast radial velocities, they do not reach the 100\kms shown at the tips of the fastest explosive filaments. These characteristics suggest a different nature and ejection mechanism of the two kind of outflows. In addition, we have shown indications that the bipolar outflow may comprise ejections in different directions, as it have been seen in Cepheus\,A\,HW2 \citep[e.g.,][]{2009_Cunningham}. In the case of Cepheus, the close interactions of a very eccentric non-coplanar companion, is thought to tilt the axis of ejection, hence driving the outflow in different directions. A similar interpretation has been claimed to explain the arches and loop structures displaying a daisy-like pattern in G5.89-0.39 \citep{2021_Fernandez-Lopez}.  If there was a multiple system comprising massive protostars at the center of the DR21 complex, we suggest that a stage characterized by gravitational interactions between its members could produce the periodic tilt of the main bipolar outflow already present before the explosion. The resulting outflowing ejections may display bow-shocks in different directions, a large range of radial velocity, and a wide opening angle bipolar appearance when seen at lower angular resolution.  Moreover, the bipolar outflow prior to the explosive event could also explain the lack of explosive streamers in the east-west direction in DR21, since most of the material from the original envelope and parental cloud may have been swept up beforehand. 

Regarding the explosive outflow, one of the current explanations for its production is via the disintegration of a multiple system with massive protostars or the coalescence of two of its members. It is easy to link the previous scenario with close passages between the members of a closely packaged system with a more chaotic outcome such as the one originating an explosive outflow. In this scenario, the main bipolar outflow in DR21 may be launched by one of the protostars involved in the event leading to the explosive outflow, and we are now seeing the remnants of its last stages. This outflow could stop at the same time that the explosive outflow was triggered.  The protostars that lie at the center of the HII region --including the one launching the bipolar outflow-- are not there any more. They may have been ejected (as they were in Orion BN/KL) in the chaotic process that ended up in triggering the explosive outflow.

\subsection{Outflow from MM5} \label{mm5outflow}
In addition to the main north-east/south-west bipolar outflow or outflows, ejected by a massive protostar, and the several explosive streamers with a common origin, associated the ALMA data reveals a red-shifted collimated outflow lobe, crossing part of the south-western lobe of the main outflow (Figure \ref{fig:bipolar_contornos_2}). This newly found outflow, runs with a positions angle of $23\deg$ and it is associated with the continuum compact source MM5. Also, since MM5 is close to the field of view edge, its blue-shifted counterpart is just hinted. This counter-lobe probably extends beyond the mosaic coverage, and may have already been imaged in HCO$^+$ at low blue-shifted velocities \citep[see channel at $-5$ to $-10$\kms in Figure 3, and integrated emission in Figure 4 of][]{2023_Skretas}. This object, comprising a 0.1-1\msun dusty disk/envelope and a classical unperturbed bipolar outflow, is possibly in the foreground/background and do not physically interact with the more powerful outflows in the DR21 region \citep[the dust mass estimate is a very rough estimate following the same assumptions as in][]{2022_Guzman}.

\section{CONCLUSIONS}  \label{sec:conclusion}

From sensitive and high angular resolution ALMA \co line observations of the DR21  massive star-forming region, we found 18 quasi-radial collimated  streamers that seem to emerge from a common center. The radial velocities along each streamer roughly follow linear gradients. Blue-shifted and red-shifted streamers are interspersed along similar directions and do not posses bipolar counterparts. Currently, there are not known protostars located at the origin of the streamers. In addition, we estimate a kinetic energy of 10$^{48}$\,erg from the ensemble of outflow streamers. All these features confirm that this outflow is likely an explosive dispersal outflow, as already suggested by \citet{2013_Zapata}. We set a new dynamical time scale for the explosive outflow of 8600\,years using an expansion velocity of 100\kms and a outflow diameter of 120\arcsec. 

Regarding the well-known infrared bipolar outflow, the high-angular resolution ALMA observations show a few indications (cavity edges at very dissimilar radial velocities, extremely high blue- and red-shifted emission at the so-called interaction region, and arcs and bow-shocks with different orientations) which suggest the presence of an outflow with multiple ejections in slightly different directions.  These bipolar ejections may explain its wide opening angle and the lack of explosive streamers with north-east/south-west orientation. We tried to reconcile these findings with the presence of the explosive outflow, apparently emanating from the same source. We speculate that both the bipolar and the explosive outflow were possibly driven by the same source. We associate the bipolar outflow with the last stages of a multiple system in disintegration, whose periodic gravitational encounters produced successive tilts in the system until it finished up with an explosive event.

Finally, the CO ALMA emission revealed still a new outflow associated with one of the five newly detected millimeter cores, MM5. These five compact cores are probably associated with disks and envelopes of young stars.

\section*{Acknowledgements}
We are very grateful to the anonymous referee for all of the
comments and suggestions that helped a lot to improve the text
and contents of this work.
E.G.C. and M.F.L. are truly grateful for all the support from UNAM's Instituto de Radioastronomía y Astrofísica of Morelia. 
M.F.L. has received funding from the European Union's
Horizon 2020 Research and Innovation Programme under the
Marie Sklodowska-Curie grant agreement No 734374 (LACEGAL).
This paper makes use of the following ALMA data: ADS/
JAO.ALMA\#019.1.00263.S. ALMA is a partnership of ESO
(representing its meMRr states), NSF (USA), and NINS
(Japan), together with NRC (Canada), NSC and ASIAA
(Taiwan), and KASI (Republic of Korea), in cooperation with
the Republic of Chile. The Joint ALMA Observatory is
operated by ESO, AUI/NRAO, and NAOJ.

Facilities: ALMA, HST.

Software: CASA (v6.4.3.27 McMullin et al. 2007), Astropy (Astropy Collaboration et al. 2013), and CARTA 2.0.0 (https://doi.org/10.5281/zenodo.4905459).

\bibliographystyle{aa}
\bibliography{biblio}

\begin{thebibliography}{46}
\expandafter\ifx\csname natexlab\endcsname\relax\def\natexlab#1{#1}\fi

\bibitem[{{Bally} {et~al.}(2022){Bally}, {Chia}, {Ginsburg}, {Reipurth}, {Tanaka}, {Zinnecker}, \& {Faulhaber}}]{2022_Bally}
{Bally}, J., {Chia}, Z., {Ginsburg}, A., {et~al.} 2022, \apj, 924, 50

\bibitem[{{Bally} {et~al.}(2011){Bally}, {Cunningham}, {Moeckel}, {Burton}, {Smith}, {Frank}, \& {Nordlund}}]{2011_Bally}
{Bally}, J., {Cunningham}, N.~J., {Moeckel}, N., {et~al.} 2011, \apj, 727, 113

\bibitem[{{Bally} {et~al.}(2017){Bally}, {Ginsburg}, {Arce}, {Eisner}, {Youngblood}, {Zapata}, \& {Zinnecker}}]{2017_Bally}
{Bally}, J., {Ginsburg}, A., {Arce}, H., {et~al.} 2017, \apj, 837, 60

\bibitem[{{Bally} {et~al.}(2020){Bally}, {Ginsburg}, {Forbrich}, \& {Vargas-Gonz{\'a}lez}}]{2020_Bally}
{Bally}, J., {Ginsburg}, A., {Forbrich}, J., \& {Vargas-Gonz{\'a}lez}, J. 2020, \apj, 889, 178

\bibitem[{{Chandler} {et~al.}(1993){Chandler}, {Gear}, \& {Chini}}]{1993_Chandler}
{Chandler}, C.~J., {Gear}, W.~K., \& {Chini}, R. 1993, \mnras, 260, 337

\bibitem[{{Cruz-Gonz{\'a}lez} {et~al.}(2007){Cruz-Gonz{\'a}lez}, {Salas}, \& {Hiriart}}]{2007_Cruz}
{Cruz-Gonz{\'a}lez}, I., {Salas}, L., \& {Hiriart}, D. 2007, \rmxaa, 43, 337

\bibitem[{{Cunningham} {et~al.}(2009){Cunningham}, {Moeckel}, \& {Bally}}]{2009_Cunningham}
{Cunningham}, N.~J., {Moeckel}, N., \& {Bally}, J. 2009, \apj, 692, 943

\bibitem[{{Cyganowski} {et~al.}(2003){Cyganowski}, {Reid}, {Fish}, \& {Ho}}]{2003_Cyganowski}
{Cyganowski}, C.~J., {Reid}, M.~J., {Fish}, V.~L., \& {Ho}, P.~T.~P. 2003, \apj, 596, 344

\bibitem[{{Davis} \& {Smith}(1996)}]{1996_Davis}
{Davis}, C.~J. \& {Smith}, M.~D. 1996, \aap, 310, 961

\bibitem[{{Dickel} {et~al.}(1978){Dickel}, {Dickel}, \& {Wilson}}]{1978_Dickel}
{Dickel}, J.~R., {Dickel}, H.~R., \& {Wilson}, W.~J. 1978, \apj, 223, 840

\bibitem[{{Fern{\'a}ndez-L{\'o}pez} {et~al.}(2014){Fern{\'a}ndez-L{\'o}pez}, {Arce}, {Looney}, {Mundy}, {Storm}, {Teuben}, {Lee}, {Segura-Cox}, {Isella}, {Tobin}, {Rosolowsky}, {Plunkett}, {Kwon}, {Kauffmann}, {Ostriker}, {Tassis}, {Shirley}, \& {Pound}}]{2014_Fernandez}
{Fern{\'a}ndez-L{\'o}pez}, M., {Arce}, H.~G., {Looney}, L., {et~al.} 2014, \apjl, 790, L19

\bibitem[{{Fern{\'a}ndez-L{\'o}pez} {et~al.}(2023){Fern{\'a}ndez-L{\'o}pez}, {Girart}, {L{\'o}pez-V{\'a}zquez}, {Estalella}, {Busquet}, {Curiel}, \& {A{\~n}ez-L{\'o}pez}}]{2023_Fernandez}
{Fern{\'a}ndez-L{\'o}pez}, M., {Girart}, J.~M., {L{\'o}pez-V{\'a}zquez}, J.~A., {et~al.} 2023, arXiv e-prints, arXiv:2307.06178

\bibitem[{{Fern{\'a}ndez-L{\'o}pez} {et~al.}(2021){Fern{\'a}ndez-L{\'o}pez}, {Sanhueza}, {Zapata}, {Stephens}, {Hull}, {Zhang}, {Girart}, {Koch}, {Cort{\'e}s}, {Silva}, {Tatematsu}, {Nakamura}, {Guzm{\'a}n}, {Nguyen Luong}, {Guzm{\'a}n Ccolque}, {Tang}, \& {Chen}}]{2021_Fernandez-Lopez}
{Fern{\'a}ndez-L{\'o}pez}, M., {Sanhueza}, P., {Zapata}, L.~A., {et~al.} 2021, \apj, 913, 29

\bibitem[{{Garden} \& {Carlstrom}(1992)}]{1992_Garden}
{Garden}, R.~P. \& {Carlstrom}, J.~E. 1992, \apj, 392, 602

\bibitem[{{Garden} {et~al.}(1991{\natexlab{a}}){Garden}, {Geballe}, {Gatley}, \& {Nadeau}}]{1991a_Garden}
{Garden}, R.~P., {Geballe}, T.~R., {Gatley}, I., \& {Nadeau}, D. 1991{\natexlab{a}}, \apj, 366, 474

\bibitem[{{Garden} {et~al.}(1991{\natexlab{b}}){Garden}, {Hayashi}, {Gatley}, {Hasegawa}, \& {Kaifu}}]{1991b_Garden}
{Garden}, R.~P., {Hayashi}, M., {Gatley}, I., {Hasegawa}, T., \& {Kaifu}, N. 1991{\natexlab{b}}, \apj, 374, 540

\bibitem[{{G{\'o}mez} {et~al.}(2008){G{\'o}mez}, {Rodr{\'\i}guez}, {Loinard}, {Lizano}, {Allen}, {Poveda}, \& {Menten}}]{2008_Gomez}
{G{\'o}mez}, L., {Rodr{\'\i}guez}, L.~F., {Loinard}, L., {et~al.} 2008, \apj, 685, 333

\bibitem[{{G{\'o}mez} {et~al.}(2005){G{\'o}mez}, {Rodr{\'I}guez}, {Loinard}, {Lizano}, {Poveda}, \& {Allen}}]{2005_Gomez}
{G{\'o}mez}, L., {Rodr{\'I}guez}, L.~F., {Loinard}, L., {et~al.} 2005, \apj, 635, 1166

\bibitem[{{Guzm{\'a}n Ccolque} {et~al.}(2024){Guzm{\'a}n Ccolque}, {Fern{\'a}ndez L{\'o}pez}, {Vazzano}, {de Gregorio}, {Plunkett}, \& {Santamar{\'\i}a-Miranda}}]{2024GuzmanCcolque}
{Guzm{\'a}n Ccolque}, E., {Fern{\'a}ndez L{\'o}pez}, M., {Vazzano}, M.~M., {et~al.} 2024, arXiv e-prints, arXiv:2403.12841

\bibitem[{{Guzm{\'a}n Ccolque} {et~al.}(2022){Guzm{\'a}n Ccolque}, {Fern{\'a}ndez-L{\'o}pez}, {Zapata}, \& {Baug}}]{2022_Guzman}
{Guzm{\'a}n Ccolque}, E., {Fern{\'a}ndez-L{\'o}pez}, M., {Zapata}, L.~A., \& {Baug}, T. 2022, \apj, 937, 51

\bibitem[{{Harris}(1973)}]{1973_Harris}
{Harris}, S. 1973, \mnras, 162, 5P

\bibitem[{{Liechti} \& {Walmsley}(1997)}]{1997_Liechti}
{Liechti}, S. \& {Walmsley}, C.~M. 1997, \aap, 321, 625

\bibitem[{{McCaughrean} \& {Pearson}(2023)}]{2023_McCaughrean}
{McCaughrean}, M.~J. \& {Pearson}, S.~G. 2023, arXiv e-prints, arXiv:2310.03552

\bibitem[{{Navarete} {et~al.}(2019){Navarete}, {Leurini}, {Giannetti}, {Wyrowski}, {Urquhart}, {K{\"o}nig}, {Csengeri}, {G{\"u}sten}, {Damineli}, \& {Menten}}]{2019Navarete}
{Navarete}, F., {Leurini}, S., {Giannetti}, A., {et~al.} 2019, \aap, 622, A135

\bibitem[{{Navarete}(2018)}]{2018Navarete}
{Navarete}, F. D.~T. 2018, PhD thesis, University of Sao Paulo, Brazil

\bibitem[{{Plambeck} \& {Menten}(1990)}]{1990_Plambeck}
{Plambeck}, R.~L. \& {Menten}, K.~M. 1990, \apj, 364, 555

\bibitem[{{Raga} {et~al.}(2021){Raga}, {Rivera-Ortiz}, {Cant{\'o}}, {Rodr{\'\i}guez-Gonz{\'a}lez}, \& {Castellanos-Ram{\'\i}rez}}]{Raga2021}
{Raga}, A.~C., {Rivera-Ortiz}, P.~R., {Cant{\'o}}, J., {Rodr{\'\i}guez-Gonz{\'a}lez}, A., \& {Castellanos-Ram{\'\i}rez}, A. 2021, \mnras, 508, L74

\bibitem[{{Rivera-Ortiz} {et~al.}(2021){Rivera-Ortiz}, {Rodr{\'\i}guez-Gonz{\'a}lez}, {Cant{\'o}}, \& {Zapata}}]{Rivera2021}
{Rivera-Ortiz}, P.~R., {Rodr{\'\i}guez-Gonz{\'a}lez}, A., {Cant{\'o}}, J., \& {Zapata}, L.~A. 2021, \apj, 916, 56

\bibitem[{{Rivera-Ortiz} {et~al.}(2019){Rivera-Ortiz}, {Rodr{\'\i}guez-Gonz{\'a}lez}, {Hern{\'a}ndez-Mart{\'\i}nez}, {Cant{\'o}}, \& {Zapata}}]{Rivera2019}
{Rivera-Ortiz}, P.~R., {Rodr{\'\i}guez-Gonz{\'a}lez}, A., {Hern{\'a}ndez-Mart{\'\i}nez}, L., {Cant{\'o}}, J., \& {Zapata}, L.~A. 2019, \apj, 885, 104

\bibitem[{{Rodr{\'\i}guez} {et~al.}(2005){Rodr{\'\i}guez}, {Poveda}, {Lizano}, \& {Allen}}]{2005_Rodriguez}
{Rodr{\'\i}guez}, L.~F., {Poveda}, A., {Lizano}, S., \& {Allen}, C. 2005, \apjl, 627, L65

\bibitem[{{Rodr{\'\i}guez-Gonz{\'a}lez} {et~al.}(2023){Rodr{\'\i}guez-Gonz{\'a}lez}, {Rivera-Ortiz}, {Castellanos-Ram{\'\i}rez}, {Raga}, {Hern{\'a}ndez-Mart{\'\i}nez}, {Cant{\'o}}, {Zapata}, \& {Robles-Valdez}}]{Rodriguez2023}
{Rodr{\'\i}guez-Gonz{\'a}lez}, A., {Rivera-Ortiz}, P.~R., {Castellanos-Ram{\'\i}rez}, A., {et~al.} 2023, \mnras, 519, 4818

\bibitem[{{Roelfsema} {et~al.}(1989){Roelfsema}, {Goss}, \& {Geballe}}]{1989_Roelfsema}
{Roelfsema}, P.~R., {Goss}, W.~M., \& {Geballe}, T.~R. 1989, \aap, 222, 247

\bibitem[{{Russell} {et~al.}(1992){Russell}, {Bally}, {Padman}, \& {Hills}}]{1992_Russell}
{Russell}, A.~P.~G., {Bally}, J., {Padman}, R., \& {Hills}, R.~E. 1992, \apj, 387, 219

\bibitem[{{Rygl} {et~al.}(2012){Rygl}, {Brunthaler}, {Sanna}, {Menten}, {Reid}, {van Langevelde}, {Honma}, {Torstensson}, \& {Fujisawa}}]{2012_Rygl}
{Rygl}, K.~L.~J., {Brunthaler}, A., {Sanna}, A., {et~al.} 2012, \aap, 539, A79

\bibitem[{{Sault} {et~al.}(1995){Sault}, {Teuben}, \& {Wright}}]{1995_Sault}
{Sault}, R.~J., {Teuben}, P.~J., \& {Wright}, M.~C.~H. 1995, in Astronomical Society of the Pacific Conference Series, Vol.~77, Astronomical Data Analysis Software and Systems IV, ed. R.~A. {Shaw}, H.~E. {Payne}, \& J.~J.~E. {Hayes}, 433

\bibitem[{{Skretas} {et~al.}(2023){Skretas}, {Karska}, {Wyrowski}, {Menten}, {Beuther}, {Ginsburg}, {Hern{\'a}ndez-G{\'o}mez}, {Gieser}, {Li}, {Kim}, {Semenov}, {Bouscasse}, {Christensen}, {Winters}, \& {Hacar}}]{2023_Skretas}
{Skretas}, I.~M., {Karska}, A., {Wyrowski}, F., {et~al.} 2023, arXiv e-prints, arXiv:2309.09687

\bibitem[{{Smith} {et~al.}(1998){Smith}, {Eisloffel}, \& {Davis}}]{1998_Smith}
{Smith}, M.~D., {Eisloffel}, J., \& {Davis}, C.~J. 1998, \mnras, 297, 687

\bibitem[{{Vazzano} {et~al.}(2021){Vazzano}, {Fern{\'a}ndez-L{\'o}pez}, {Plunkett}, {de Gregorio-Monsalvo}, {Santamar{\'\i}a-Miranda}, {Takahashi}, \& {Lopez}}]{2021_Vazzano}
{Vazzano}, M.~M., {Fern{\'a}ndez-L{\'o}pez}, M., {Plunkett}, A., {et~al.} 2021, \aap, 648, A41

\bibitem[{{Wu} {et~al.}(2004){Wu}, {Wei}, {Zhao}, {Shi}, {Yu}, {Qin}, \& {Huang}}]{2004_Wu}
{Wu}, Y., {Wei}, Y., {Zhao}, M., {et~al.} 2004, \aap, 426, 503

\bibitem[{{Yamaguchi} {et~al.}(1999){Yamaguchi}, {Akira}, \& {Yasuo}}]{1999_Yamaguchi}
{Yamaguchi}, R., {Akira}, M., \& {Yasuo}, F. 1999, in Star Formation 1999, ed. T.~{Nakamoto}, 383--384

\bibitem[{{Zapata} {et~al.}(2013{\natexlab{a}}){Zapata}, {Fernandez-Lopez}, {Curiel}, {Patel}, \& {Rodriguez}}]{2013Zapata}
{Zapata}, L.~A., {Fernandez-Lopez}, M., {Curiel}, S., {Patel}, N., \& {Rodriguez}, L.~F. 2013{\natexlab{a}}, arXiv e-prints, arXiv:1305.4084

\bibitem[{{Zapata} {et~al.}(2023){Zapata}, {Fern{\'a}ndez-L{\'o}pez}, {Leurini}, {Guzm{\'a}n Ccolque}, {Rodriguez}, {Palau}, {Menten}, \& {Wyrowski}}]{2023_Zapata}
{Zapata}, L.~A., {Fern{\'a}ndez-L{\'o}pez}, M., {Leurini}, S., {et~al.} 2023, arXiv e-prints, arXiv:2309.11386

\bibitem[{{Zapata} {et~al.}(2020){Zapata}, {Ho}, {Fern{\'a}ndez-L{\'o}pez}, {Ccolque}, {Rodr{\'\i}guez}, {Reyes-Vald{\'e}s}, {Bally}, {Palau}, {Saito}, {Sanhueza}, {Rivera-Ortiz}, \& {Rodr{\'\i}guez-Gonz{\'a}lez}}]{2020_Zapata}
{Zapata}, L.~A., {Ho}, P. T.~P., {Fern{\'a}ndez-L{\'o}pez}, M., {et~al.} 2020, \apjl, 902, L47

\bibitem[{{Zapata} {et~al.}(2009){Zapata}, {Schmid-Burgk}, {Ho}, {Rodr{\'\i}guez}, \& {Menten}}]{2009_Zapata}
{Zapata}, L.~A., {Schmid-Burgk}, J., {Ho}, P. T.~P., {Rodr{\'\i}guez}, L.~F., \& {Menten}, K.~M. 2009, \apjl, 704, L45

\bibitem[{{Zapata} {et~al.}(2013{\natexlab{b}}){Zapata}, {Schmid-Burgk}, {P{\'e}rez-Goytia}, {Ho}, {Rodr{\'\i}guez}, {Loinard}, \& {Cruz-Gonz{\'a}lez}}]{2013_Zapata}
{Zapata}, L.~A., {Schmid-Burgk}, J., {P{\'e}rez-Goytia}, N., {et~al.} 2013{\natexlab{b}}, \apjl, 765, L29

\bibitem[{{Zapata} {et~al.}(2017){Zapata}, {Schmid-Burgk}, {Rodr{\'\i}guez}, {Palau}, \& {Loinard}}]{2017_Zapata}
{Zapata}, L.~A., {Schmid-Burgk}, J., {Rodr{\'\i}guez}, L.~F., {Palau}, A., \& {Loinard}, L. 2017, \apj, 836, 133

\end{thebibliography}


\begin{appendix}
\section{Outflow mass estimate using the emission of CO}\label{appx:masses}

We estimate the mass in each velocity channel as: 

\begin{equation}
    M=\mu mh\Omega N_{tot}/X_{CO}      
\end{equation}

where $\mu$ is the mean molecular weight, which is assumed to be equal 2.76 \citep{1999_Yamaguchi}, \textit{m} is the hydrogen atom mass ($\sim 1.67\times10^{-24}g$), h is the Planck's constant ($6.6261\times10^{-27}\,erg\,s$), $\Omega$ is the solid angle of the emitting region, and a CO abundance of X$_{CO}$=10$^{-4}$. For to derive the column density of the \co transition we use the equation: 

$$N_{tot}(CO)=\frac{3h}{8 \pi^{3} \mu^{2}_{B} J} \frac{Q_{rot}\, e^{E_{u}/kT_{ex}}}{e^{h\nu/kT_{ex}}-1} \frac{\int T_B dv}{[J_{\nu}(T_{ex})-J_{\nu}(T_{bg})]}= $$

\begin{equation}
    =\frac{1.195 \times 10^{14} (T_{ex}+0.922) e^{16.596/T_{ex}}}{e^{11.065/T_{ex}}-1} \frac{T_{B} \Delta v}{J_{\nu}(T_{ex})-J_{\nu}(T_{bg})}     
\end{equation}

In the expression above we use the Planck's constant $h=6.6261\times10^{-27}\,erg\,s$, the dipole moment $\mu_{B}=1.1011\times10^{-19}$\,StatC\,cm, the quantum number of the upper level $J=2$, the partition function $Q_{rot}=kT_{ex}/(hB_{0})+1/3$ with a rigid rotor rotation constant $B_{0}=57.635968$\,GHz, the Boltzmann's constant $k=1.3806\time10^{-16}\,erg\,k^{-1}$, the Rayleight-Jeans equivalent temperature $J_{\nu}(T)=(h\nu/k)/(e^{h\nu/kT}-1)$, the brightness temperature $T_{exc}$ and $T_{B}$ in K and the channel width $\Delta v$ in\kms.

\section{Moment maps of CO streamers}\label{appx:cubos}
\renewcommand\thefigure{B.\arabic{figure}}    
\setcounter{figure}{0}  

\begin{figure*}
    \centering
    \includegraphics[scale=0.5]{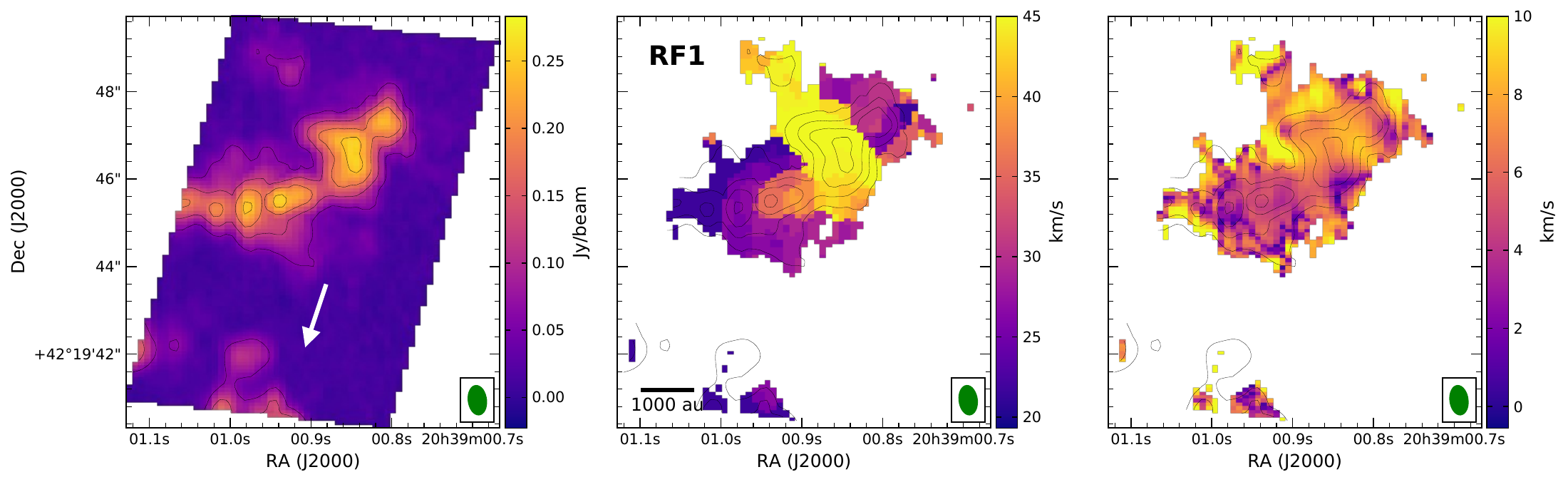}
    \includegraphics[scale=0.5]{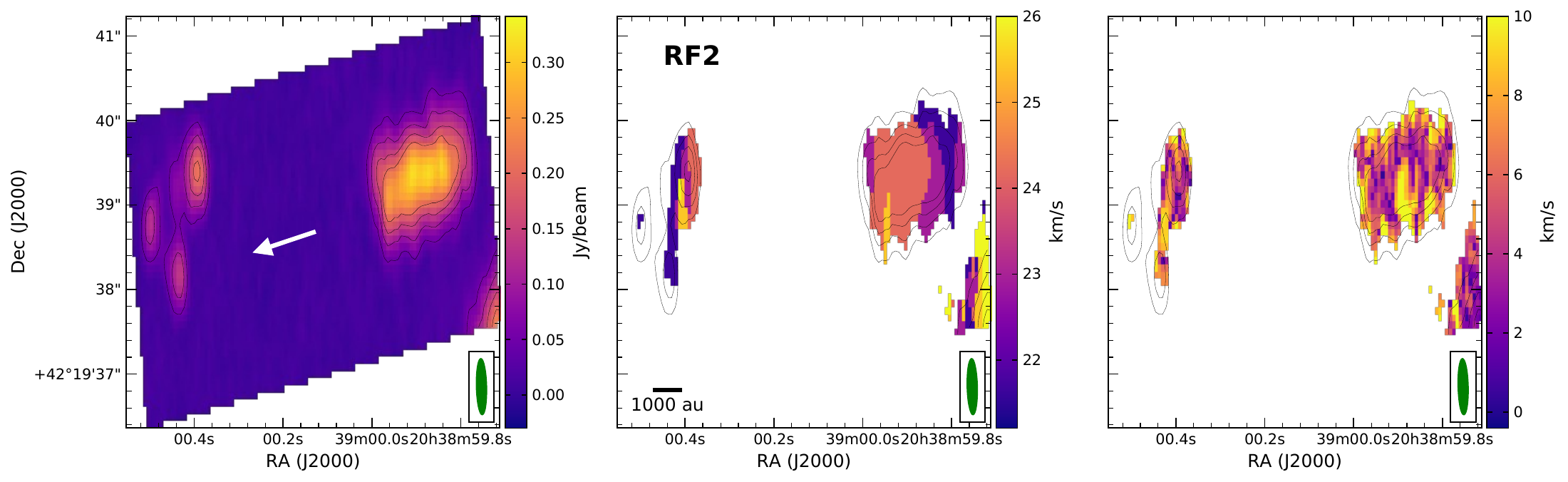}
    \includegraphics[scale=0.5]{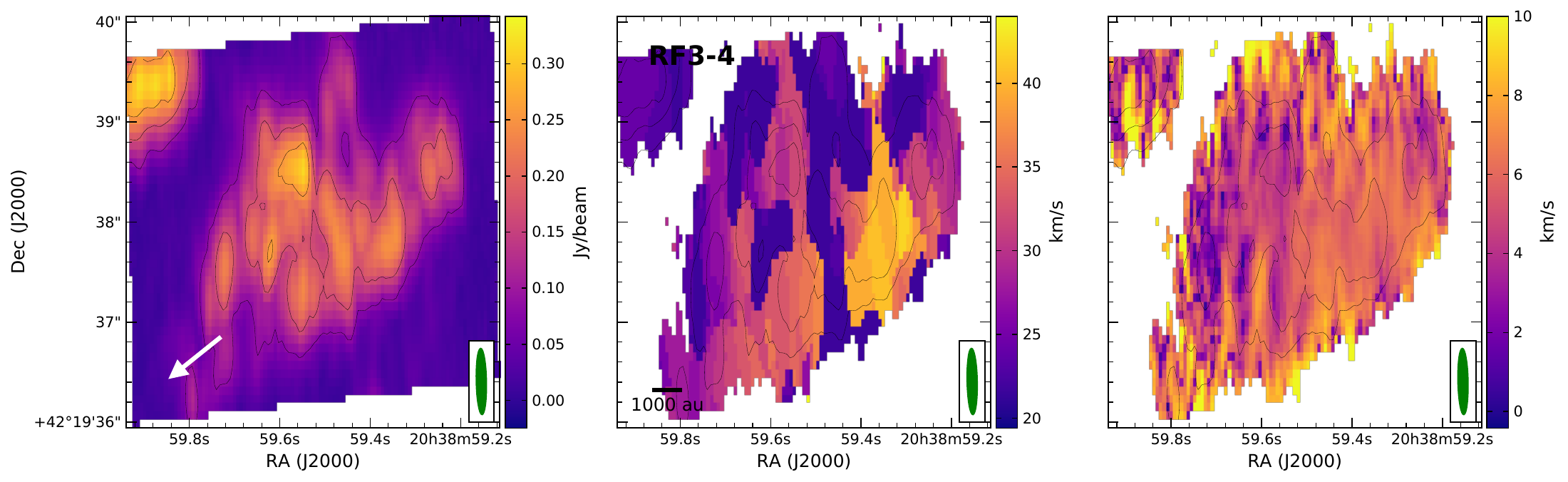}
    \caption{Moment-8,moment-9 and moment-2 maps (from left to right) of the \co redshifted streamers of DR21: RF1, RF2, RF3, RF4 and RF5. The black contours represents the \co emission at 5, 10, 15 and 20 times the rms noise level for each moment-8 map. The synthesized beam of the \co map is shown within a box at the bottom right corner. The white arrow indicates the direction of the center of the explosive outflow.} 
    \label{mom1_red}
\end{figure*}

\begin{figure*}
    \centering 
    \includegraphics[scale=0.5]{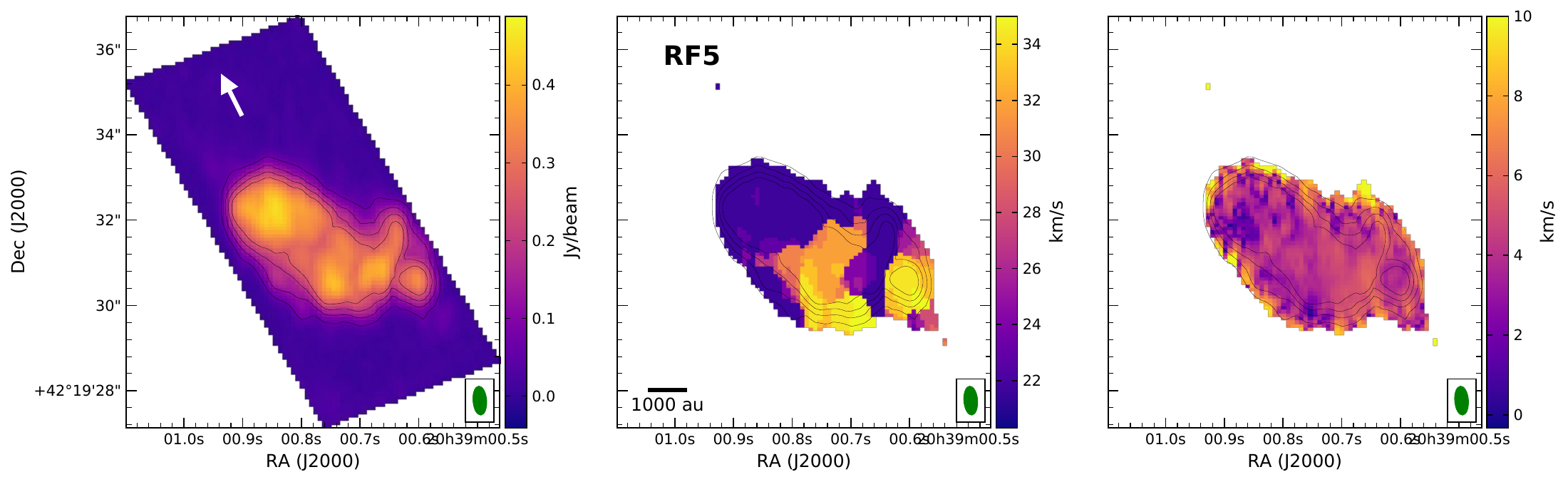}
    \includegraphics[scale=0.5]{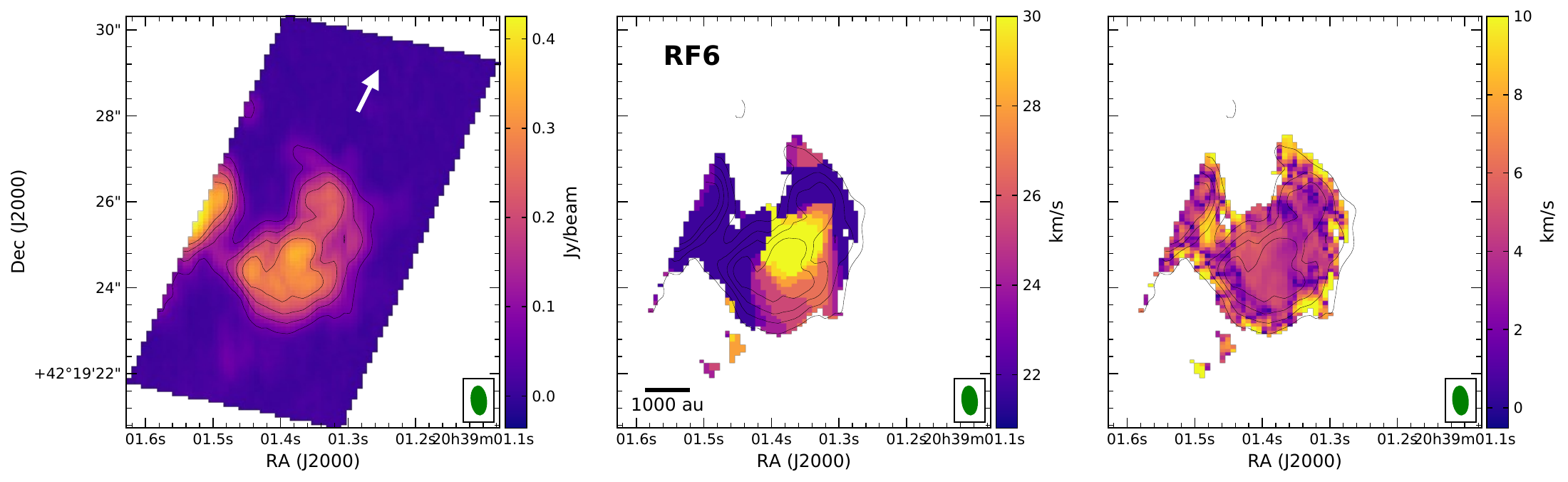} 
    \includegraphics[scale=0.5]{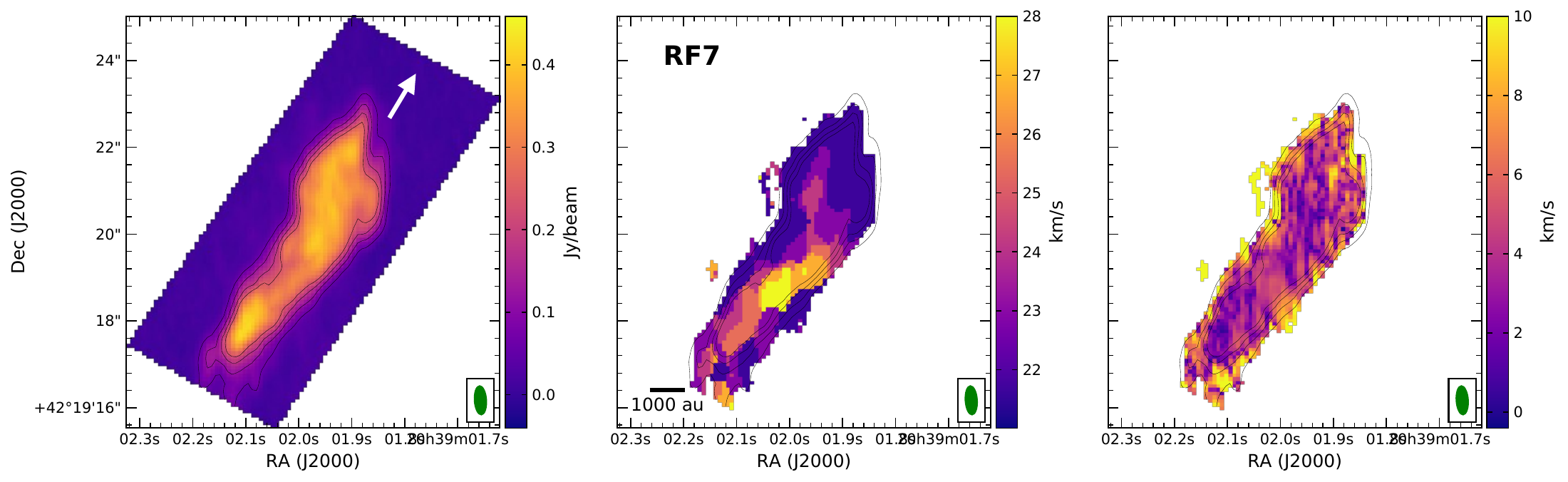}
    \includegraphics[scale=0.5]{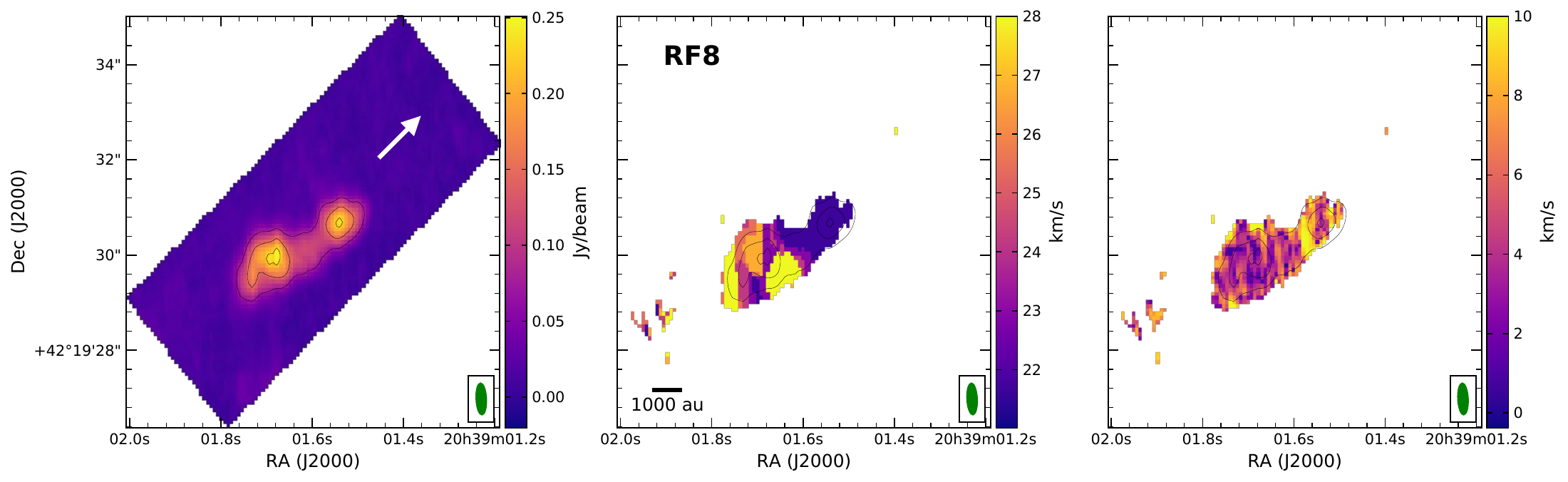}
    \caption{Same as Fig. \ref{mom1_red} for redshifted streamers RF6, RF7, RF8 and RF9.} 
    \label{mom2_red}
\end{figure*}

\begin{figure*}
    \centering
   \includegraphics[scale=0.5]{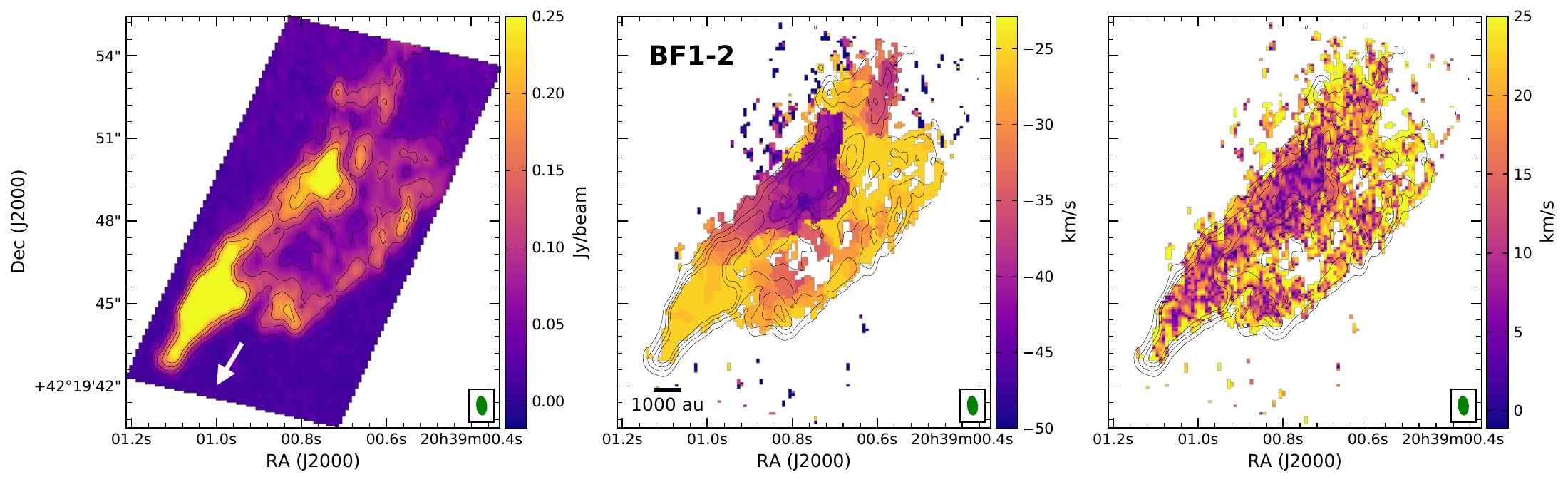}
    \includegraphics[scale=0.5]{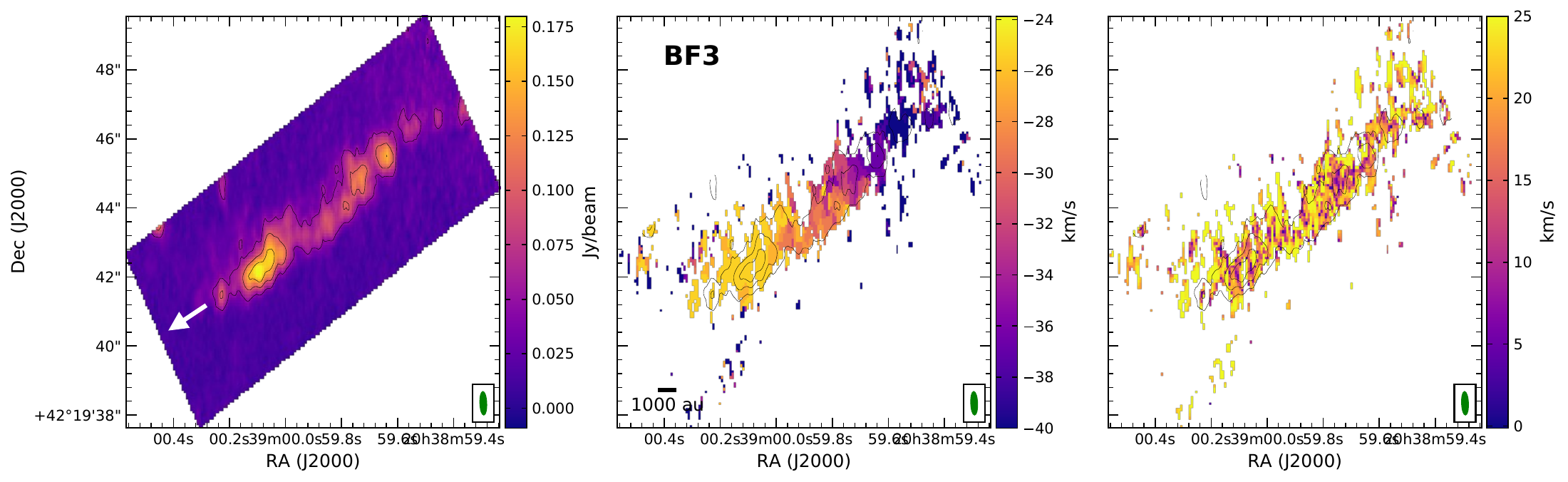}
    \includegraphics[scale=0.5]{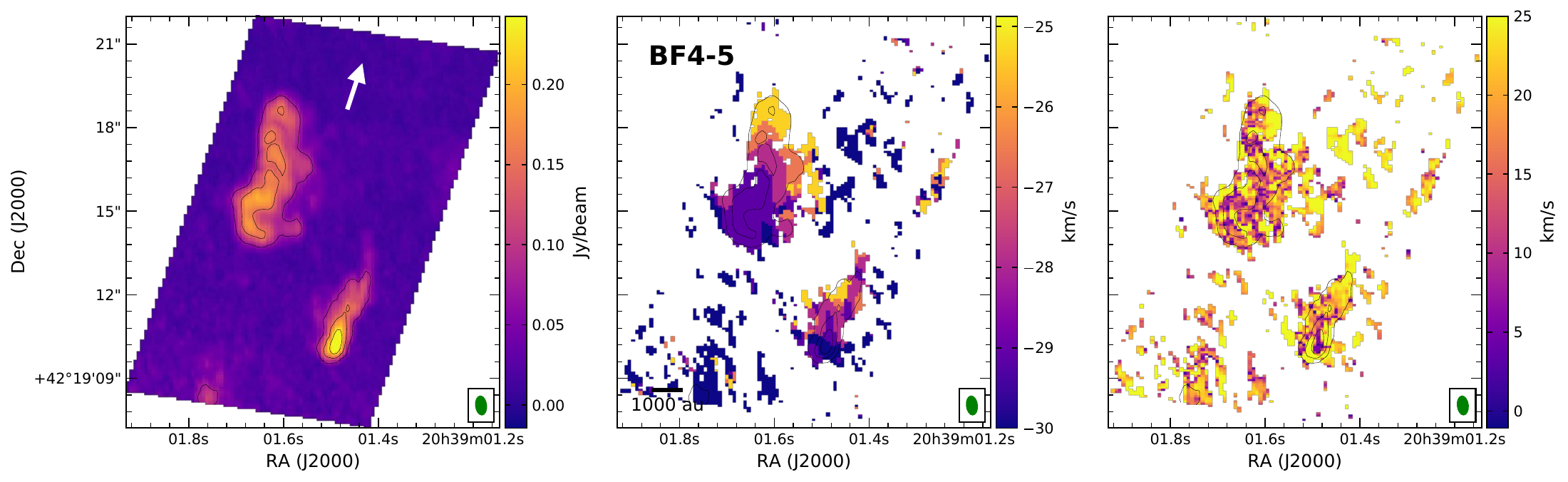}
    \includegraphics[scale=0.5]{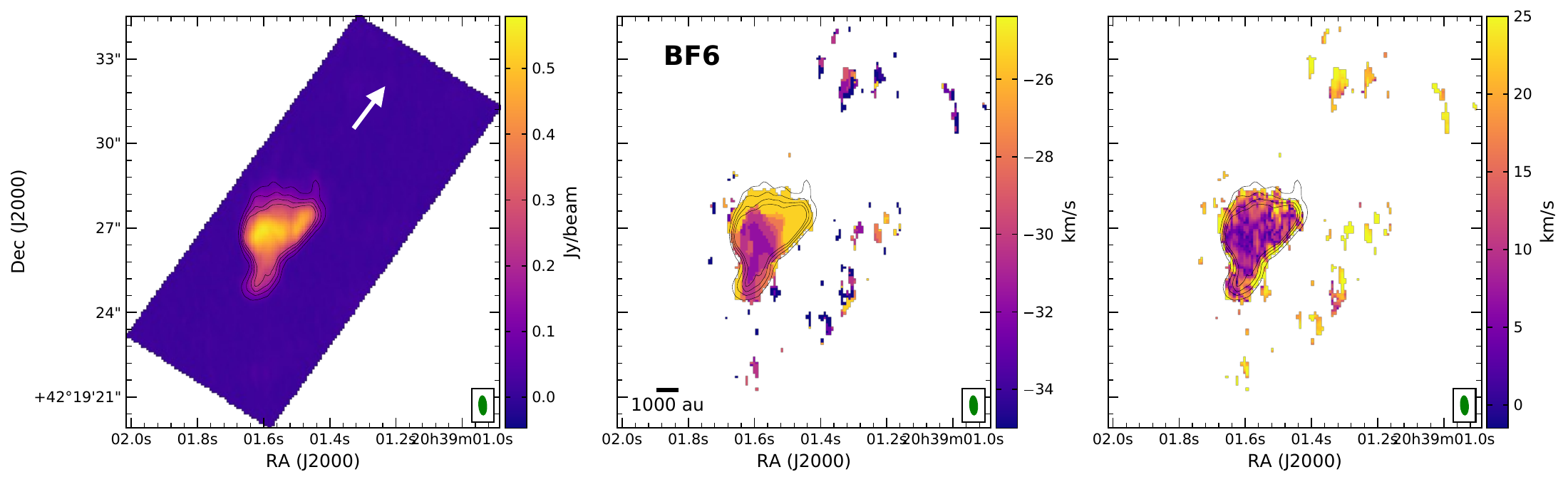}
    \caption{Same as Fig. \ref{mom1_red} for blueshifted streamers BF1, BF2, BF3, BF4, BF5 and BF6.} 
    \label{mom1_blue}
\end{figure*}

\begin{figure*}
    \centering
    \includegraphics[scale=0.5]{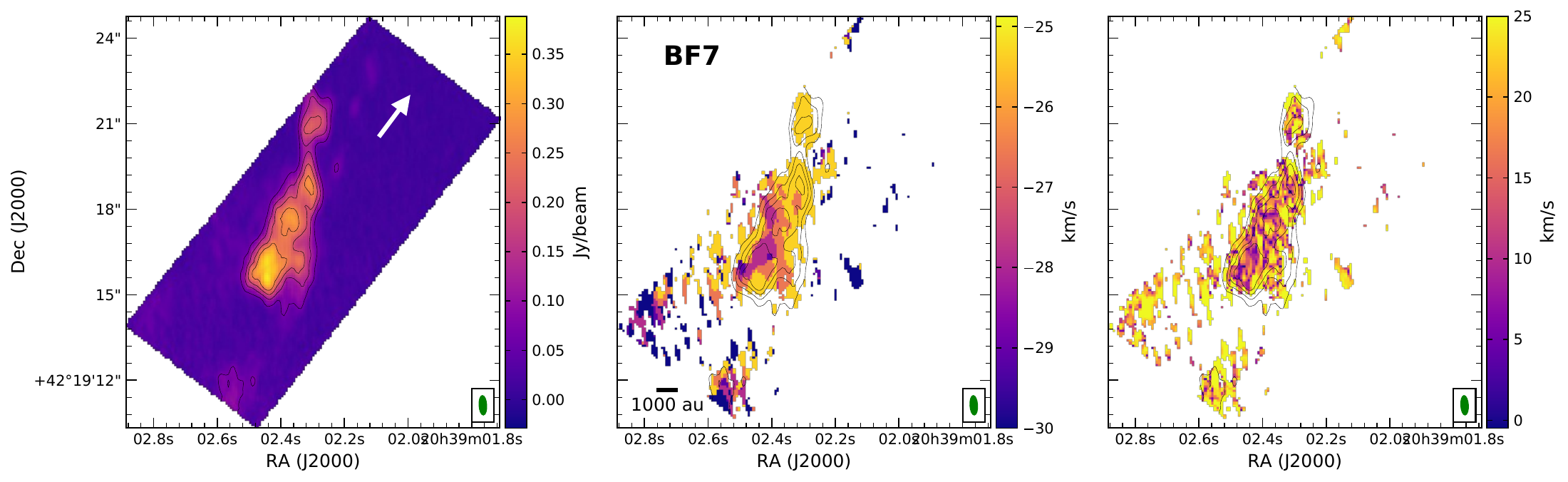}   
    \includegraphics[scale=0.5]{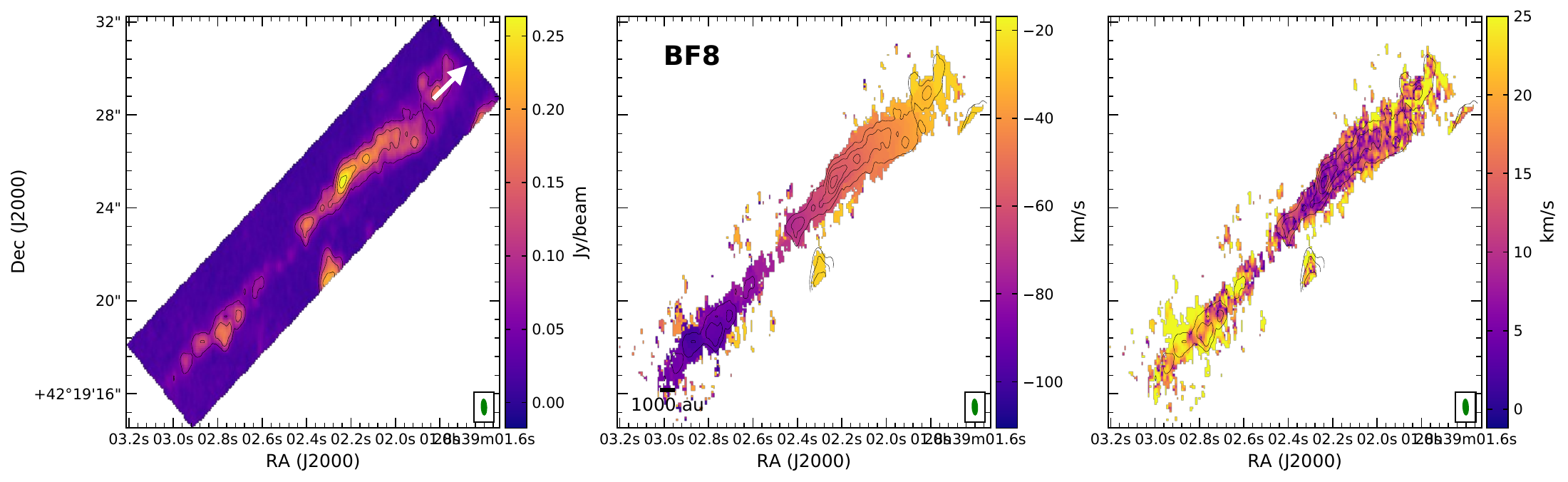}
    \includegraphics[scale=0.5]{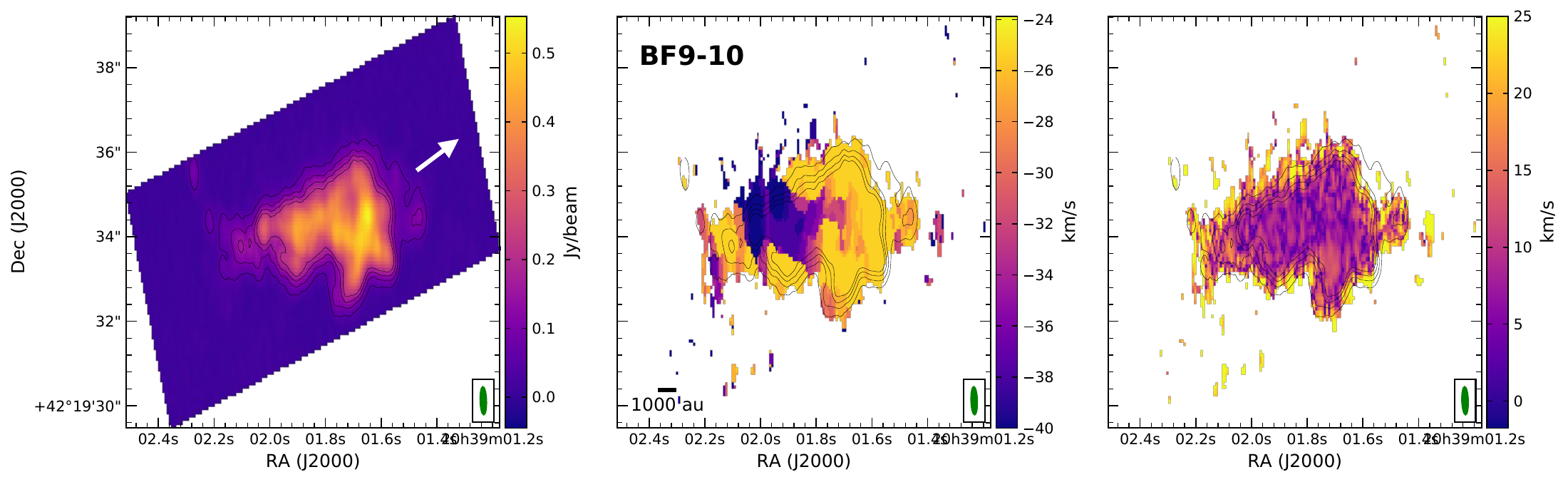}
    \caption{Same as Fig. \ref{mom1_red} for blueshifted streamers BF7, BF8, BF9 and BF10.} 
    \label{mom2_blue}
\end{figure*}

Figures \ref{mom1_red}-\ref{mom2_blue} show the moment-8, moment-9 and moment-2 maps (from left to right) of \co line emission for each red-shifted and blue-shifted streamers of DR21 identified in this paper.  \\

\end{appendix}


\end{document}